\documentclass[pre,showpacs,showkeys,preprintnumbers,amsmath,amssymb,superscriptaddress,twocolumn]{revtex4-2}
\pdfoutput=1 
\usepackage{graphicx}
\usepackage{amssymb,amsfonts,amsmath}
\usepackage{epstopdf}
\usepackage{color}
\usepackage{bm}
\usepackage[colorlinks=true,linkcolor=blue,citecolor=red]{hyperref}
\usepackage{comment}
\usepackage{multirow}

\def\pa{\partial\Omega}

\def\R{{\mathbb R}}

\def\ctanh{\mathrm{ctanh}}

\def\x{\bm{x}}

\def\pa{{\partial \Omega}}

\def\R{\mathbb{R}}

\def\Z{\mathbb{Z}}

\def\A{\mathbf{A}}

\def\MM{\mathbf{M}}
\def\mm{\mathbf{P}}
\def\V{\mathbf{V}}

\def\nmax{{n_{\rm max}}}

\begin{document}

\title{Imperfect diffusion-controlled reactions on a torus and on a pair of balls} 

\author{Denis~S.~Grebenkov}
 \email{denis.grebenkov@polytechnique.edu}

\affiliation{Laboratoire de Physique de la Mati\`{e}re Condens\'{e}e (UMR 7643), \\ 
CNRS -- Ecole Polytechnique, Institut Polytechnique de Paris, 91120 Palaiseau, France}

\date{Received: \today / Revised version: }

\begin{abstract}
We employ a general spectral approach based on the Steklov eigenbasis
to describe imperfect diffusion-controlled reactions on bounded
reactive targets in three dimensions.  The steady-state concentration
and the total diffusive flux onto the target are expressed in terms of
the eigenvalues and eigenfunctions of the exterior Steklov problem.
In particular, the eigenvalues are shown to provide the geometric
lengthscales of the target that are relevant for diffusion-controlled
reactions.  Using toroidal and bispherical coordinates, we propose an
efficient procedure for analyzing and solving numerically this
spectral problem for an arbitrary torus and a pair of balls,
respectively.  A simple two-term approximation for the diffusive flux
is established and validated.  Implications of these results in the
context of chemical physics and beyond are discussed.
\end{abstract}



\maketitle

\section{Introduction}
\label{sec:intro}

Diffusion-controlled reactions play an important role in various
natural phenomena and chemical industry
\cite{House,Murrey,Lindenberg,Grebenkov}.  In his seminal work,
Smoluchowski unveiled the importance of diffusion as a mixing
mechanism for reactant molecules and established the first
mathematical framework to describe physical, chemical and biological
processes, in which the diffusive transport is the rate-limiting step
\cite{Smoluchowski18}.  While his original work was focused on the
coagulation kinetics, the proposed framework found numerous
applications in heterogeneous catalysis, recombination reactions,
etc. (see reviews
\cite{North66,Wilemski73,Calef83,Berg85,Rice85,Weiss86,Szabo89,Zhou10,Grebenkov23}
and references therein).  In a nutshell, the Smoluchowski's approach
describes multiple point-like species $A$ that diffuse independently
from each other with a constant diffusion coefficient $D$ and
instantly react upon hitting an immobile catalytic sphere $B$ of
radius $R$: $A + B \to B$.  Such an immediate reaction ensures zero
concentration of species $A$ on the perfectly reactive target.
Solving the diffusion equation with a prescribed concentration $C_A$
at infinity, Smoluchowski determined the time-dependent concentration
of species $A$, as well as their diffusive flux $J$ onto the target,
which can be interpreted as the overall reaction rate.  In the
steady-state regime, his famous relation reads
\begin{equation} \label{eq:JSmol}
J = 4\pi DC_A R, 
\end{equation}
in which the flux is proportional to the size of the target, not to
its surface area.  This is a signature of diffusion-limited reactions.

The respective roles of the geometric shape of the target and of its
reactivity have been thoroughly investigated.  On one hand, Collins
and Kimball introduced the notion of partial reactivity of the target
by replacing the Dirichlet boundary condition, $C_A = 0$, by a more
general Robin boundary condition,
\begin{equation}  \label{eq:Robin}
-D \partial_n C_A = \kappa \, C_A \quad \textrm{on the target},
\end{equation}
where $\partial_n = (\bm{n} \cdot \nabla)$ is the normal derivative
oriented outwards the bulk (i.e., inwards the target)
\cite{Collins49}.  This condition postulates that the diffusive flux
of species $A$ (left-hand side) is proportional to the reactive flux
(right-hand side), and the proportionality coefficient $\kappa$
characterizes the reactivity of the catalytic target $B$, ranging from
$0$ for inert surface, to infinity for a perfectly reactive one (in
which case it is reduced to the Smoluchowski setting).  The reactivity
$\kappa$ is often related to the reaction constant $k_{\rm on}$ and
may represent microscopic heterogeneity of the target, temporal
fluctuations of its activity, energetic or entropic barriers,
etc. \cite{Sano79,Brownstein79,Sano81,Shoup82,Zwanzig90,Sapoval94,Powles92,Berezhkovskii04,Traytak07,Bressloff08,Singer08b,Lawley15,Guerin21,Piazza22,Guerin23,Cengiz24}.
Microscopic and probabilistic interpretations of the condition
(\ref{eq:Robin}) have been discussed \cite{Grebenkov20}.

For a spherical target of radius $R$, Collins and Kimball employed the
separation of variables and found an explicit form of the
time-dependent concentration, as well as the diffusive flux.  In the
steady-state regime, the diffusive flux reaches a constant value:
\begin{equation}  \label{eq:J_sphere}
J  = \frac{4\pi D C_A R}{1 + D/(\kappa R)} \,.
\end{equation}
For weakly reactive targets ($\kappa \ll D/R$), the flux $J \approx
4\pi R^2 \kappa C_A$ is proportional to the surface area $4\pi R^2$ of
the target and to its reactivity $\kappa$, and is almost independent
of diffusivity $D$; this is the reaction-limited regime.  In the
opposite limit of highly reactive targets ($\kappa \gg D/R$), one
retrieves the Smoluchowski relation (\ref{eq:JSmol}).  When the shape
of the target is nearly spherical, one can employ perturbative
analysis to generalize this relation \cite{Piazza19}.  It is also
instructive to look at the ratio between the imposed initial
concentration $C_A$ (a ``voltage'') and the diffusive flux $J$ (a
``current'') that represents the ``resistance'' (or impedance) of the
system:
\begin{equation}  \label{eq:Zsphere}
Z = \frac{C_A}{J} = \frac{1}{4\pi RD} + \frac{1}{4\pi R^2\kappa} \,.  
\end{equation}
In analogy to electrostatics, this relation is often interpreted as
the sum of ``resistances'' of the bulk diffusion and of the surface
reaction, as two consecutive steps to the successful reaction event.
However, this intuitive interpretation relies on the simple form
(\ref{eq:J_sphere}), which is valid for the spherical target but fails
for arbitrary domains (see below).

On the other hand, the Smoluchowski relation (\ref{eq:JSmol}) remains
valid for a perfectly reactive target of arbitrary shape if $R$ is
replaced by its capacitance $R_c$:  
\begin{equation}  \label{eq:JSmol_general}
J = 4\pi D C_A R_c
\end{equation}
(here we use the convention that the capacitance of a sphere of radius
$R$ is $R$).  For several basic domains whose symmetries allow for
separation of variables, analytical computation of the capacitance
relies on curvilinear orthogonal coordinates \cite{Smythe}.  So,
Samson and Deutch pioneered the application of bispherical coordinates
to compute the diffusive flux on a pair of perfectly reactive
spherical targets \cite{Samson77}.  Later many authors used the
bispherical coordinates to obtain analytical solutions for
diffusion-controlled reactions on a pair of spherical targets and to
model surface reactions on axially symmetric Janus dumbbell particles
\cite{Zoia98,Strieder00,Piazza05,Bluett06,Popescu11,Kapral15,Michelin17}.
Numerous studies were devoted to understand the impact of the spatial
arrangment of multiple perfectly reactive targets onto
diffusion-controlled reactions by using Monte Carlo simulations
\cite{Zheng89,Tsao01,Kansal02} or finite-element methods
\cite{Eun13}, as well as semi-analytical approaches such as the
generalized method of separation of variables
\cite{Galanti16,Galanti16b,Traytak18,Grebenkov19f,Grebenkov20f}.

\begin{figure}
\begin{center}
\includegraphics[width=85mm]{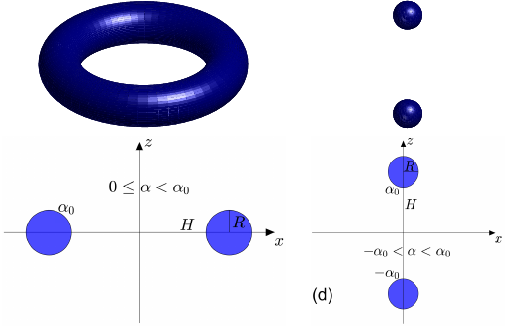}
\end{center}
\caption{
{\bf (Left)} Illustration of a torus with major radius $H = 1$ and
minor radius $R = 0.25$ (top) and its projection onto $xz$ plane
(bottom).  {\bf (Right)} Illustration of a pair of identical balls of
radius radius $R = 0.25$, whose centers are separated by distance $2H$
(top) and its projection onto $xz$ plane (bottom).}
\label{fig:torus}
\end{figure}

In this paper, we aim at clarifying the respective roles of shape and
reactivity onto the diffusive flux onto nonspherical targets.  For
this purpose, we adopt a spectral approach, which was originally
developed for electrochemical settings with irregularly-shaped
electrodes \cite{Grebenkov06b,Filoche08} and then extended to
time-dependent diffusion \cite{Grebenkov20}.  In particular, we
explain the failure of an oversimplified picture of consecutive
diffusion and reaction resistances that was inspired by
Eq. (\ref{eq:Zsphere}) for a sphere, and get a general spectral
representation for the diffusive flux.  We then focus on two cases, a
torus and a pair of balls in $\R^3$ (Fig. \ref{fig:torus}), for which
the proposed spectral description can be efficiently computed by using
toroidal and bispherical coordinates.  In particular, we derive a
simple two-term approximation for the diffusive flux.  Section
\ref{sec:main} presents our main results and their implications in the
context of chemical physics, whereas technical derivations are
relegated to Appendices \ref{sec:torus} and \ref{sec:bisphere}.

\section{Main results}
\label{sec:main}

In this Section, we present our main theoretical results for the
diffusive flux onto a partially reactive target in three dimensions.
The target is a compact set $\Omega_0$ whose boundary $\pa$ is
reactive but impenetrable for the species $A$, which diffuse in the
bulk $\Omega = \R^3 \backslash \Omega_0$ with a constant diffusion
coefficient $D > 0$.  Upon the arrival onto the catalytic surface
$\pa$ of reactivity $\kappa > 0$, the species $A$ may react with a
finite probability or resume its bulk diffusion.  The concentration
gradient is therefore established, and we search for a steady-state
concentration of $A$, satisfying
\begin{equation}  \label{eq:uq_def}
\left\{ \begin{array}{r l l}
\Delta C_A & = 0 & \textrm{in}~\Omega, \\ - D\partial_n C_A & = \kappa C_A & \textrm{on}~ \pa, \\
C_A(\x) & \to C_A & \textrm{as}~ |\x|\to \infty, \\ \end{array}\right.
\end{equation}
where $C_A$ is a prescribed concentration at infinity, and $\Delta$ is
the Laplace operator.  The total diffusive flux is 
\begin{equation}
J = \int\limits_{\pa} (-D\partial_n C_A).
\end{equation}
We aim at determining the dependence of $J$ on the reactivity $\kappa$
and the shape of the boundary $\pa$.

\subsection{General spectral expansions}

To get a spectral representation for a solution of the problem
(\ref{eq:uq_def}), we employ the eigenvalues $\mu_k$ and
eigenfunctions $V_k$ of the exterior Steklov problem:
\begin{equation} \label{eq:Steklov}
\left\{ \begin{array}{r l l} 
\Delta V_k & = 0  & \textrm{in}~\Omega, \\
\partial_n V_k & = \mu_k V_k  & \textrm{on}~ \pa, \\
V_k &\to 0  & \textrm{as}~|\x|\to\infty , \\ \end{array} \right.
\end{equation}
with the conventional normalization
\begin{equation}
\int\limits_{\pa} |V_k|^2 = 1.
\end{equation}
Different mathematical formulations of the exterior Steklov problem
and their equivalence are discussed in
\cite{Auchmuty14,Arendt15,Bundrock25} (see also
\cite{Levitin,Girouard17,Colbois24} for surveys on a much more
well-studied Steklov problem for bounded domains).  Note that a
different formulation of the spectral problem (\ref{eq:Steklov}) in
the context of sloshing phenomena in hydrodynamics appeared in
\cite{Henrici70,Troesch72,Miles72,Fox83,Kozlov04}.
In sharp contrast to the continuous spectrum of the Laplace operator
in unbounded domains, the spectrum of the Steklov problem
(\ref{eq:Steklov}) is discrete because the boundary $\pa$ is bounded;
the eigenvalues can be ordered in an increasing sequence,
\begin{equation}
0 < \mu_0 \leq \mu_1 \leq \cdots \nearrow +\infty ,
\end{equation}
whereas the eigenfunctions $V_k$, restricted to $\pa$, form a complete
orthonormal basis of $L^2(\pa)$.  This peculiar mathematical property
makes the Steklov eigenmodes particularly suitable for solving
exterior problems
\cite{Auchmuty04,Auchmuty13,Auchmuty15,Auchmuty18} and for describing
diffusion-controlled reactions in unbounded domains
\cite{Grebenkov21a,Grebenkov25}.  Note that restrictions of $V_k$ onto
$\pa$, $(V_k)|_{\pa}$, are the eigenfunctions of the associated
Dirichlet-to-Neumann operator.

Writing $C_A(\x) = C_A (1 - u(\x))$ and using the completeness of the
eigenfunctions $(V_k)|_{\pa}$ in $L^2(\pa)$ to represent the unknown
function $u(\x)$, one can deduce the following spectral representation
(see \cite{Grebenkov20} for more details):
\begin{equation}  \label{eq:u_spectral}
C_A(\x) = C_A \biggl(1 - \sum\limits_{k=0}^\infty \frac{V_k(\x)}{1 + \mu_k D/\kappa} \int\limits_{\pa} d\x \, V_k(\x)\biggr).
\end{equation}
It also determines the diffusive flux onto the target:
\begin{equation}  \label{eq:J_spectral}
J = D C_A |\pa|\sum\limits_{k=0}^\infty \frac{F_k}{\mu_k^{-1} + D/\kappa} \,,
\end{equation}
where $|\pa|$ is the surface area of the target, and
\begin{equation}  \label{eq:Fk}
F_k = \frac{1}{|\pa|}\left|\int\limits_{\pa} d\x \, V_k(\x)\right|^2 
\end{equation}
are dimensionless coefficients.  For a spherical target, the Steklov
spectrum is known explicitly (see, e.g., \cite{Grebenkov20}); in
particular, all coefficients $F_k$ are zero, except for $F_0 = 1$ so
that Eq. (\ref{eq:J_spectral}) is reduced to Eq. (\ref{eq:J_sphere}),
with $\mu_0^{-1} = R$.  However, this simplification is a consequence
of the rotational symmetry of the spherical target, and it does not
hold in general.  In fact, the spectral representation
(\ref{eq:J_spectral}) indicates that the impact of a finite reactivity
$\kappa$ onto the diffusive flux is in general much more sophisticated
than that for a spherical target.  In particular, an intuitive
interpretation of the reaction event via Eq. (\ref{eq:Zsphere}) as a
sequence of the diffusion and reaction steps is not true anymore.  In
fact, the first arrival of the particle onto the target is generally
followed by a sequence of failed reactions attempts and diffusive
excursions in the bulk near the target, until the successful reaction
event \cite{Grebenkov20}.  In this way, the geometric structure of the
target, which is represented via the spectral parameters $\mu_k$ and
$F_k$, is coupled to the reactivity $\kappa$.

A similar spectral representation for the electric current between two
metallic electrodes was derived in \cite{Grebenkov06b}.  Despite an
evident similarity between two problems, the geometric setting in
\cite{Grebenkov06b} and thus the employed Steklov problem were
different.  In fact, Ref. \cite{Grebenkov06b} aimed at representing
microscopic irregularity of a macroscopically large electrode by
repeating periodically a given geometric pattern of the electrode
surface in both lateral directions.  This is mathematically equivalent
to considering the current between an infinitely large (unbounded)
metallic surface and a source at infinity.  In turn, the surface $\pa$
of a target in our setting is bounded.  Even though this distinction
may sound as a technical detail, it drastically changes the spectral
properties of the Steklov problem.  For instance, the principal
eigenvalue $\mu_0$ was zero for the periodic setting of
Ref. \cite{Grebenkov06b}, whereas it is strictly positive in our case.
In turn, the coefficients $F_k$ in \cite{Grebenkov06b} involved
projections of $V_k$ onto the harmonic measure density, in contrast to
our Eq. (\ref{eq:Fk}).  Many other spectral properties are different
as well.  At the same time, most physical interpretations from
\cite{Grebenkov06b,Filoche08} remain valid in our setting.

It is instructive to inspect the limits of low and high reactivity.
As $\kappa\to 0$, one can employ the completeness relation for the
eigenfunctions $(V_k)|_{\pa}$ in $L^2(\pa)$ to show that
\begin{equation}  \label{eq:sumFk}
\sum\limits_{k=0}^\infty F_k = 1,
\end{equation}
and thus to retrieve the expected reaction-limited rate:
\begin{equation}
J \approx C_A \kappa |\pa|   \qquad (\kappa \to 0).
\end{equation}
In this low-reactivity regime, the whole boundary of the target is
accessible almost uniformly by diffusion, so that $J$ is proportional
to the surface area $|\pa|$ and to the reactivity $\kappa$ but is
almost independent of the diffusion coefficient $D$.  Note also that
Eq. (\ref{eq:sumFk}) allows one to interpret the coefficients $F_k$
are relative weights of different Steklov eigenmodes in the spectral
expansion (\ref{eq:J_spectral}).  In turn, the Steklov eigenvalues
$\mu_k$ determine the set of characteristic lengthscales,
$\mu_k^{-1}$, of the target.  These geometric lengthscales are then
confronted to the ``reactivity length'', $D/\kappa$, which was
introduced and thoroughly discussed in
\cite{Sapoval94,Sapoval02,Felici03,Felici05,Grebenkov15}.

In the opposite limit of high reactivity, one recovers the generalized
Smoluchowski relation (\ref{eq:JSmol_general}), with the capacitance
\begin{equation}   \label{eq:Capacitance}
R_c = \frac{|\Omega|}{4\pi} \sum\limits_{k=0}^\infty \mu_k F_k .
\end{equation}
One sees that Eq. (\ref{eq:J_spectral}) describes a smooth transition
between these two limits.  In particular, the relevant geometric
information of the target is fully captured via the spectral
parameters $\mu_k$ and $F_k$.  In order to clarify the role of these
spectral parameters onto diffusion-controlled reactions, we consider
two surfaces of revolution: a torus and a pair of spheres.  The use of
toroidal and bispherical coordinates will allow us to design a very
efficient and accurate method for computing the Steklov eigenmodes.
In the next two sections, we present the main results for these
targets, whereas most technical computations are relegated to
Appendices \ref{sec:torus} and \ref{sec:bisphere}.

\subsection{Torus}

We consider diffusion in the exterior of a torus with the major radius
$H$ and the minor radius $R$: $\Omega = \{ (x,y,z)\in \R^3~:~
(\sqrt{x^2+y^2} - H)^2 + z^2 > R^2\}$ (Fig. \ref{fig:torus}).  In
Appendix \ref{sec:torus}, we describe the computation of the
eigenvalues, eigenfunctions, and spectral weights $F_k$ for the
associated exterior Steklov problem.  The fast convergence of our
method allows to deal with truncated matrices of small size (we use
the truncation order $\nmax = 10$ in the following examples).  Note
that Eqs. (\ref{eq:sumFk}, \ref{eq:Capacitance}) can serve for
checking the accuracy of the numerical computation.  We recall that
the capacitance of the torus has an exact explicit expression
\cite{Belevitch83},
\begin{equation}  \label{eq:Rc_torus}
R_c = \frac{4}{\pi} \sqrt{H^2-R^2} \sum\limits_{n=0}^\infty 
\frac{Q_{n-\frac12}(H/R)}{(1+\delta_{n,0})P_{n-\frac12}(H/R)} \,,
\end{equation}
where $P_\nu(z)$ and $Q_\nu(z)$ are Legendre functions of the first
and second kind, and $\delta_{n,0}$ is the Kronecker symbol (see
Appendix \ref{sec:torus_Rc} for details).

\begin{table}
\begin{center}
\begin{tabular}{|c|c|c|c|c|c||c|c|c|c|c|} \hline
     & \multicolumn{5}{|c||}{Torus} & \multicolumn{5}{|c|}{Pair of balls} \\ \hline
$R/H$& $F_0$ & $F_1$  & $F_2$ &$\mu_0$&$\mu_1$&$F_0$ & $F_1$   & $F_2$   & $\mu_0$ & $\mu_1$ \\  \hline
0.99 & 0.49 & 0.45    & 0.05    & 0.47  & 0.58  & 0.79 & 0.19    & 0.01    & 0.58 & 1.12 \\ 
0.9  & 0.67 & 0.32    & 0.01    & 0.47  & 0.76  & 0.93 & 0.07    &$10^{-4}$& 0.72 & 1.69 \\
0.8  & 0.76 & 0.23    &$10^{-3}$& 0.50  & 0.94  & 0.96 & 0.04    &$10^{-4}$& 0.86 & 2.11 \\
0.7  & 0.83 & 0.17    &$10^{-3}$& 0.54  & 1.15  & 0.98 & 0.02    &$10^{-4}$& 1.04 & 2.57 \\
0.6  & 0.88 & 0.12    &$10^{-3}$& 0.60  & 1.41  & 0.99 & 0.01    &$10^{-4}$& 1.27 & 3.12 \\
0.5  & 0.92 & 0.08    &$10^{-4}$& 0.68  & 1.77  & 0.99 & 0.01    &$10^{-4}$& 1.59 & 3.85 \\
0.4  & 0.95 & 0.05    &$10^{-4}$& 0.80  & 2.29  & 1.00 &$10^{-3}$&$10^{-5}$& 2.08 & 4.90 \\
0.3  & 0.97 & 0.03    &$10^{-5}$& 0.99  & 3.16  & 1.00 &$10^{-3}$&$10^{-5}$& 2.90 & 6.61 \\
0.2  & 0.99 & 0.01    &$10^{-6}$& 1.33  & 4.86  & 1.00 &$10^{-4}$&$10^{-6}$& 4.54 & 9.97 \\ 
0.1  & 1.00 &$10^{-3}$&$10^{-8}$& 2.27  & 9.91  & 1.00 &$10^{-5}$&$10^{-8}$& 9.52 & 19.99 \\ \hline
\end{tabular}
\end{center}
\caption{
First three coefficients $F_k$ and the first two eigenvalues $\mu_k$
(multiplied by $H = 1$) as functions of the aspect ratio $R/H$, for
the torus with major radius $H$ and minor radius $R$, and for a pair
of equal balls of radius $R$, whose centers are separated by $2H$ (see
Appendices \ref{sec:torus} and \ref{sec:bisphere} for details).}
\label{tab:muk}
\end{table}

Table \ref{tab:muk} presents the first three coefficients $F_k$ as
functions of the aspect ratio $R/H$.  One sees that the first two
eigenmodes provide the major contribution to the diffusive flux,
whereas the contribution of the third and higher eigenmode is
negligible.  This observation suggests a two-term approximation:
\begin{equation}  \label{eq:J_approx}
J^{\rm app} = D C_A |\pa|\biggl(\frac{F_0}{\mu_0^{-1} + D/\kappa} + \frac{F_1}{\mu_1^{-1} + D/\kappa}\biggr) ,
\end{equation}
where $|\pa| = 4\pi^2 RH$, and the spectral parameters $\mu_k$ and
$F_k$ as functions of $R/H$ are given in Table \ref{tab:muk} (see
Appendix \ref{sec:torus} for their numerical computation).  Moreover,
as the torus gets thinner (i.e., $R/H$ decreases), the contribution of
the principal eigenmode rapidly approaches $1$, whereas all other
eigenmodes can be neglected.  We get thus a fully explicit
approximation for a thin torus (a ring)
\begin{equation}  \label{eq:J_asympt}
J \approx  \frac{4\pi^2 D C_A H}{\ln (8H/R) + D/(\kappa R)}  \qquad (R \ll H),
\end{equation}
where we used the asymptotic relation (\ref{eq:mu_ext_torus}) for the
principal eigenvalue $\mu_0$.  In the limit $\kappa \to \infty$, we
retrieve the generalized Smoluchowski relation
(\ref{eq:JSmol_general}), with the known approximate expression for
the capacitance of a thin torus: $R_c \approx \pi H/\ln(8H/R)$.  As
for a spherical target in Eq. (\ref{eq:Zsphere}), the ``impedance''
$C_A/J$ of a thin torus can be represented as a serial connection of
``diffusion resistance'', $\ln(8H/R)/(4\pi^2 DH)$, and ``reaction
resistance'', $1/(\kappa |\pa|)$.  Interestingly, the decrease of the
minor radius $R$ has a marginal effect onto the first term (as
$\ln(H/R)$), but significantly enhances the second term (as $H/R$).
In other words, even for a highly reactive thin torus, the reaction
resistance can be dominant, and an intuitive replacement of a large
but finite $\kappa$ by infinity may lead to erroneous conclusions.
This observation highlights the crucial role of the reactivity, which
was often ignored in former works.  Note that a similar effect of the
reactivity onto first-reaction times was discussed in
\cite{Grebenkov17a,Grebenkov18}.  

From a broader perspective, the consideration of a finite reactivity
resolves an optimal shape paradox.  In fact, one may wonder what is
the ``optimal way'' to redistribute a given amount of catalytic
material to get the largest diffusive flux (and thus the highest
production rate)?  If the catalytic germs are distributed uniformly on
a surface of the torus, one can fix the surface area, $|\pa| = 4\pi^2
RH$, and then search for the optimal ratio $R/H$ that maximizes the
diffusive flux.  When the reactivity $\kappa$ is infinite, the
diffusive flux diverges as $R/H \to 0$, i.e., it can be made
arbitrarily large by taking a large but thin torus.  In other words, a
fixed amount of catalytic material can yield an arbitrarily high
production rate.  This mathematically correct but physically
paradoxical statement originates from the assumption of infinite
reactivity.  In turn, if the reactivity is finite, the limit $R \to 0$
yields the diffusive flux $C_A |\pa| \kappa$, as in the
reaction-limited regime.  More generally, for any target shape, one
can rewrite the spectral expansion (\ref{eq:J_spectral}) as
\begin{equation}  \label{eq:J_spectral2}
J = C_A |\pa| \kappa \biggl(1 - \sum\limits_{k=0}^\infty \frac{F_k}{1 + \mu_k D/\kappa}\biggr).
\end{equation}
Since $F_k \geq 0$ and $\mu_k > 0$, the diffusive flux cannot exceed
its upper bound $C_A |\pa| \kappa$ for any fixed surface area $|\pa|$
and finite reactivity $\kappa$.

\begin{figure}
\begin{center}
\includegraphics[width=88mm]{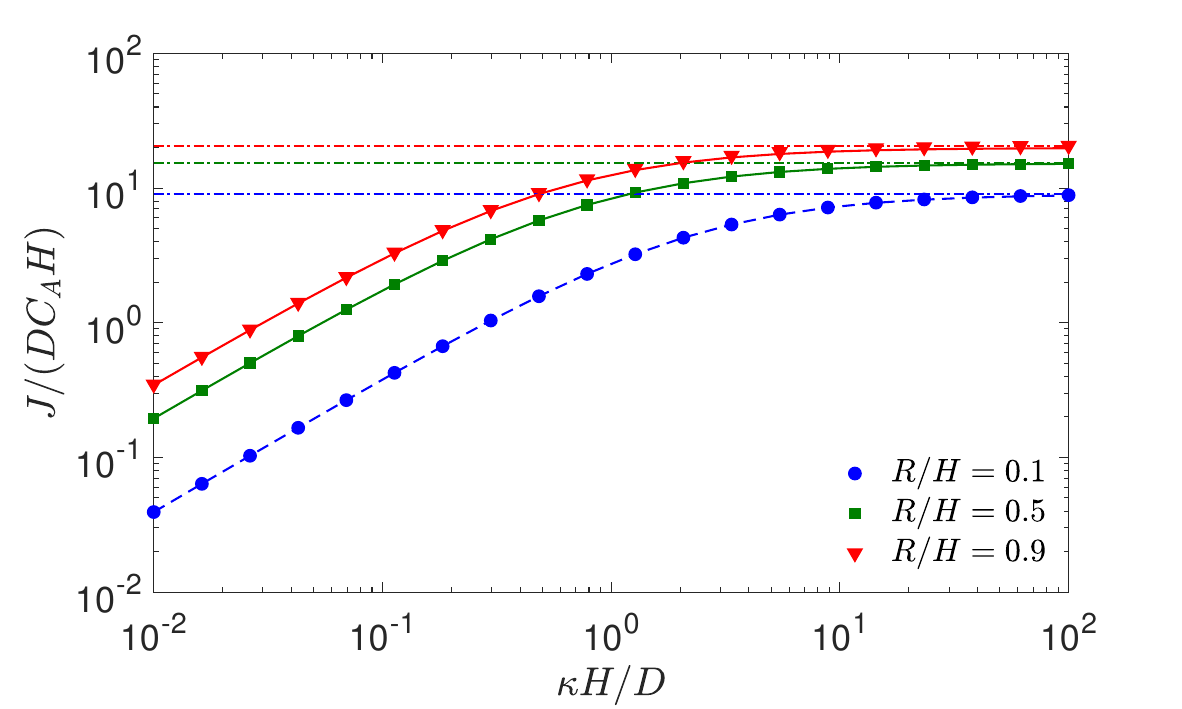} 
\end{center}
\caption{
The diffusive flux (rescaled by $DC_A H$) onto the torus with the
fixed major radius $H$ and three minor radii $R$ (see the legend), as
a function of $\kappa H/D$.  Symbols present the spectral expansion
(\ref{eq:J_spectral}), referred to as the exact solution.  Solid lines
indicate its two-term approximation (\ref{eq:J_approx}), dashed line
shows the asymptotic relation (\ref{eq:J_asympt}) for the case $R/H =
0.1$, whereas dash-dotted lines illustrate the limiting values
(\ref{eq:JSmol_general}), with the capacitance given by
Eq. (\ref{eq:Rc_torus}).}
\label{fig:Fk_torus}
\end{figure}

Figure \ref{fig:Fk_torus} presents the diffusive flux as a function of
the reactivity $\kappa$ for three tori.  Expectedly, the flux
monotonously grows from $0$ at $\kappa = 0$ to the maximal value for a
perfectly reactive target ($\kappa = \infty$), which is determined by
the capacitance in Eq. (\ref{eq:JSmol_general}).  The two-term
approximation (\ref{eq:J_approx}), shown by solid lines, is in
excellent agreement with the exact solution given by the full spectral
expansion (\ref{eq:J_spectral}).  Moreover, for the torus with $R/H =
0.1$, dashed line presents the asymptotic relation
(\ref{eq:J_asympt}), which is also in a perfect agreement with the
exact solution over the whole range of $\kappa$.  The two-term
approximation (\ref{eq:J_approx}) and the asymptotic relation
(\ref{eq:J_asympt}) present one of the main practical results of this
paper for the torus.

\subsection{Pair of equal spheres}

As discussed in Sec. \ref{sec:intro}, diffusion-limited reactions on a
pair of perfectly reactive spheres were thoroughly investigated in the
past.  In their pioneering work, Samson and Deutch obtained the
capacitance \cite{Samson77}
\begin{equation}  \label{eq:Samson}
R_c = 2R \sum\limits_{n=1}^\infty (-1)^{n-1} \frac{\sinh(\alpha_0)}{\sinh(n\alpha_0)}  
\end{equation}
that determines the diffusive flux via Eq. (\ref{eq:JSmol_general}),
where $\alpha_0 = \ln(H/R + \sqrt{H^2/R^2-1})$.  This is the simplest
setting to analyze the effect of diffusional screening
\cite{Sapoval94}, also known as diffusive interactions
\cite{Traytak92}, when two (or many) sinks compete for diffusing
particles so that the total flux is smaller than the sum of diffusive
fluxes onto each sink, if the other were absent.  This effect, which
plays an important role in physics, electrochemistry, and biology
\cite{Sapoval02,Sapoval02,Felici03,Felici05}, e.g., in
diffusion-limited growth processes
\cite{Witten81,Vicsek,Banks,Krapivsky}, was thoroughly studied for
small targets by many authors (see
\cite{Calef83,Berg85,Rice85,Weiss86,Szabo89,Zhou10} and references
therein).  Its extension to partially reactive targets was also
discussed \cite{Grebenkov19f,Grebenkov20f}.

The advantage of the geometric model of two balls is the possibility
of using bispherical coordinates to get either analytical, or rapidly
converging numerical representations.  In Appendix \ref{sec:bisphere},
we establish an efficient procedure for computing and analyzing the
eigenvalues and eigenfunctions of the exterior Steklov problem for a
pair of spheres.  This spectral information gives access to the
steady-state concentration and to the diffusive flux and its
dependence on the reactivity $\kappa$.  Even though the general case
of two spheres of arbitrary radii was considered, we focus here on the
simpler setting of a pair of equal spheres.

\begin{figure}
\begin{center}
\includegraphics[width=88mm]{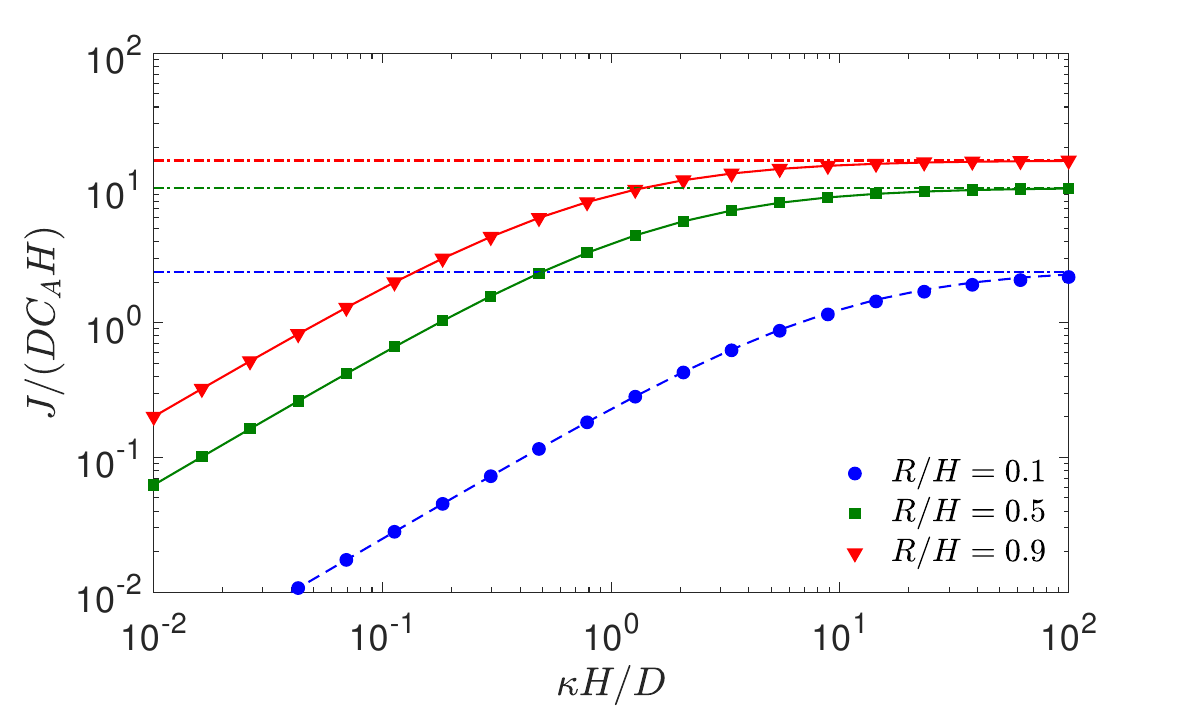} 
\end{center}
\caption{
The diffusive flux (rescaled by $DC_A H$) onto a pair of equal spheres
of radii $R$ (see the legend), whose centers are separated by a fixed
distance $2H$, as a function of $\kappa H/D$.  Symbols present the
spectral expansion (\ref{eq:J_spectral}), referred to as the exact
solution.  Solid lines indicate its two-term approximation
(\ref{eq:J_approx}), dashed line shows the asymptotic relation
(\ref{eq:J_asympt2}) for the case $R/H = 0.1$, whereas dash-dotted
lines illustrate the limiting values (\ref{eq:JSmol_general}), with
the capacitance given by Eq. (\ref{eq:Samson}).}
\label{fig:Jq_twospheres}
\end{figure}

Table \ref{tab:muk} and Figure \ref{fig:Jq_twospheres} summarize the
main results that are relevant in the context of chemical physics and
resemble those obtained for a torus.  Indeed, the first two eigenmodes
provide the dominant contribution to the diffusive flux so that the
two-term approximation (\ref{eq:J_approx}) is also valid for a pair of
spheres.  The high accuracy of this approximation is confirmed
numerically (compare symbols and solid lines in
Fig. \ref{fig:Jq_twospheres}).  Moreover, when $R \ll H$, the
principal eigenmode is dominant, yielding the fully explicit
asymptotic result:
\begin{equation}  \label{eq:J_asympt2}
J \approx \frac{8\pi R^2 D C_A}{R + D/\kappa} \,.
\end{equation}
One sees that this approximation is very accurate already for $R/H =
0.1$.  This is also consistent with the small-target approximation
discussed in \cite{Chaigneau22}.

\section{Conclusion}

In summary, we revisited the classical problem of steady-state
diffusion towards a reactive target, with a special emphasis onto the
respective roles of the shape and reactivity.  Employing the spectral
approach based on the exterior Steklov problem, we obtained a general
representation (\ref{eq:J_spectral}) of the diffusive flux onto the
target that exhibits an explicit dependence on the reactivity and
reveals the impact of the shape via the spectral parameters $\mu_k$
and $F_k$ -- the Steklov eigenvalues and the squared projections of
the associated eigenfunctions onto a constant.  This is the only
geometric information that fully determines the diffusive flux and
that can thus potentially be accessed via experimental measurements
\cite{Filoche08}.  In particular, $\mu_k^{-1}$ can be interpreted as
relevant lengthscales of the target that have to be compared to the
reaction length $D/\kappa$; in turn, $F_k$ are the relative weights of
different eigenmodes.  This spectral representation indicates that a
common view of diffusion and reaction steps as serially connected
resistances is in general wrong.  In fact, the successful reaction
event is preceded by a long sequence of failed reaction attempts, so
that diffusion and reaction steps are strongly coupled.  

For the cases of a torus and a pair of balls, we proposed an efficient
procedure for solving numerically the exterior Steklov problem.  We
obtained that two eigenmodes provide the dominant contribution to the
diffusive flux, yielding an accurate two-term approximation
(\ref{eq:J_approx}).  The four parameters $H\mu_0, H\mu_1, F_0, F_1$
depend only on the aspect ratio $R/H$, and this dependence can be
found numerically or asymptotically (see Table \ref{tab:muk}).  Using
the solution of the exterior Steklov problem for prolate and oblate
spheroids \cite{Grebenkov24}, we also computed the weights $F_k$ for
such targets and checked that the two-term approximation is very
accurate as well (not shown).  In addition, the principal eigenmode
yields a fully explicit and accurate approximation for the diffusive
flux onto a thin torus, as well as onto a pair of small balls.  While
the second setting is rather conventional and well studied, the case
of a torus does not seem to be reported earlier.

The presented theoretical results shed light onto the intricate
coupling between geometry and reactivity in diffusion-controlled
reactions.  Moreover, the knowledge of the Steklov eigenbasis allows
one to deal with more sophisticated surface reactions, far beyond the
conventional setting of a constant reactivity.  On one hand, the
Steklov eigenfunctions can be used to incorporate spatially
heterogeneous reactivity via a suitable matrix representation
\cite{Grebenkov19}.  On the other hand, one can implement an 
encounter-dependent reactivity to model passivation or activation of
the reactive target via its encounters with diffusing particles
\cite{Grebenkov20}.  Indeed, the Steklov eigenbasis is the cornerstone
of the encounter-based approach to diffusion-controlled reactions
\cite{Grebenkov20c,Grebenkov22a,Bressloff22d,Bressloff22b}
and to permeation processes
\cite{Bressloff22a,Bressloff22c,Bressloff23a,Bressloff23b}.
Finally, one can go beyond the context of chemical physics and
describe the statistics of encounters between the diffusing particles
and the target \cite{Grebenkov19c,Grebenkov21a,Grebenkov22d}.

\begin{acknowledgments}
The author thanks A. Chaigneau for complementary numerical tests with
a finite-element method \cite{Grebenkov25}.  The author acknowledges
the Simons Foundation for supporting his sabbatical sojourn in 2024 at
the CRM (CNRS -- University of Montr\'eal, Canada), as well as the
Alexander von Humboldt Foundation for support within a Bessel Prize
award.
\end{acknowledgments}

\section*{Data Availability Statement}

The data that support the findings of this study are available from
the corresponding author upon reasonable request.

\appendix

\section{Torus}
\label{sec:torus}

In this Appendix, we describe the main steps for computing the
eigenvalues and eigenfunctions of the exterior Steklov problem for a
torus with the major radius $H$ and the minor radius $R$: $\Omega = \{
(x,y,z)\in \R^3~:~ (\sqrt{x^2+y^2} - H)^2 + z^2 > R^2\}$
(Fig. \ref{fig:torus}).
For this purpose, we use the toroidal coordinates
$(\alpha,\theta,\phi)$ \cite{Morse} 
%
\begin{equation}
\left(\begin{array}{c} x \\ y \\ z \\ \end{array}\right) = \frac{c}{\cosh \alpha - \cos\theta}
\left(\begin{array}{c} \sinh\alpha \cos\phi \\ \sinh\alpha \sin \phi \\ \sin\theta \\ \end{array}\right) ,
\end{equation}
where $0 \leq \alpha < +\infty$, $-\pi < \theta \leq \pi$, $0 \leq
\phi < 2\pi$, and $c = \sqrt{H^2-R^2}$.  In these coordinates, the
exterior of a torus corresponds to $0 \leq
\alpha < \alpha_0$, where 
\begin{equation}  \label{eq:Rcond}
\alpha_0 = \cosh^{-1}(H/R) = \ln\biggl(H/R + \sqrt{(H/R)^2 - 1}\biggr).
\end{equation}
The scale factors are
\begin{equation}
h_\alpha = h_\theta = \frac{c}{\cosh\alpha - \cos\theta}, \quad h_\phi = \frac{c\, \sinh\alpha}{\cosh\alpha - \cos\theta} \,.
\end{equation}
The Laplace operator admits the partial separation of variables in
terms of associated Legendre functions $P_{n-\frac12}^m(\cosh\alpha)$
and $Q_{n-\frac12}^m(\cosh\alpha)$, as well as Fourier harmonics in
$\theta$ and $\phi$ \cite{Morse}.  Since
$P_{n-\frac12}^m(\cosh\alpha)$ [resp., $Q_{n-\frac12}^m(\cosh\alpha)$]
exhibits a logarithmic singularity as $\alpha\to \infty$ [resp,
$\alpha \to 0$], one has to choose $P_{n-\frac12}^m(\cosh\alpha)$ for
solving exterior problems and $Q_{n-\frac12}^m(\cosh\alpha)$ for
solving interior problems.

\subsection{Exterior Steklov problem}

For the exterior problem ($0 \leq \alpha < \alpha_0$), we search a
solution of the Steklov problem (\ref{eq:Steklov}) in the form
\begin{align} \nonumber
V & = \sqrt{\cosh\alpha - \cos\theta}  \sum\limits_{m\in\Z}^\infty \sum\limits_{n=0}^\infty 
\biggl[A_{m,n}^{+} \cos(n\theta)  \\   \label{eq:auxil2}
& + A_{m,n}^{-} \sin(n\theta)\biggr] \frac{P_{n-\frac12}^{|m|}(\cosh\alpha)}{P_{n-\frac12}^{|m|}(\cosh\alpha_0)} e^{im\phi}.
\end{align}
The unknown coefficients $A_{m,n}^{\pm}$ are determined by imposing
the Steklov boundary condition at the boundary $\alpha = \alpha_0$:
\begin{align*}
& (\partial_n V)|_{\pa} = \frac{1}{h_{\alpha_0}} (\partial_\alpha V)|_{\alpha_0} 
= \frac{\sqrt{\cosh\alpha_0 - \cos\theta} \sinh\alpha_0}{2c} \\
& \quad \times \sum\limits_{m,n}  \bigl[A_{m,n}^{+} \cos(n\theta) + A_{m,n}^{-} \sin(n\theta)\bigr] e^{im\phi} \\
& \quad \times \biggl\{ 1 + 2(\cosh\alpha_0-\cos\theta) p_{m,n} \biggr\} \\
& = \mu V|_{\alpha_0} = \mu \sqrt{\cosh\alpha_0 - \cos\theta}  \\
& \quad \times \sum\limits_{m,n} \bigl[A_{m,n}^{+} \cos(n\theta) + A_{m,n}^{-} \sin(n\theta)\bigr] e^{im\phi},
\end{align*}
where we used the shortcut notation $\sum\nolimits_{m,n}$ for the sum
over $m\in\Z$ and $n = 0,1,2,\ldots$, and
\begin{equation}  \label{eq:pmn_def}
p_{m,n} = p_{-m,n} = \frac{P_{n-\frac12}^{m'}(\cosh\alpha_0)}{P_{n-\frac12}^{m}(\cosh\alpha_0)} \qquad (m = 0,1,2,\ldots),
\end{equation}
with prime denoting the derivative with respect to the argument (i.e.,
$P_\nu^{\mu'}(z) = \partial_z P_\nu^m(z)$).  Using the recurrence
relation
\begin{equation}
\partial_z P_\nu^\mu(z) = \frac{1}{z^2-1} \biggl(\sqrt{z^2-1} P_\nu^{\mu+1}(z) + \mu z P_\nu^\mu(z)\biggr),
\end{equation}
we can express $p_{m,n}$ as
\begin{equation}  \label{eq:pmn_def2}  
p_{m,n} = \frac{1}{\sinh \alpha_0} \left( \frac{P_{n-\frac12}^{m+1}(\cosh\alpha_0)}{P_{n-\frac12}^{m}(\cosh\alpha_0)} 
+ m \, \ctanh(\alpha_0)\right).
\end{equation}
We rewrite then the above condition more compactly as
\begin{align} \nonumber
& \sum\limits_{m',n'}  \bigl[A_{m',n'}^{+} \cos(n'\theta) + A_{m',n'}^{-} \sin(n'\theta)\bigr] e^{im'\phi}  \\  \label{eq:auxil1}
& \times \bigl\{ 1 + 2(\cosh\alpha_0-\cos\theta) p_{m',n'} - 2R \mu \bigr\} = 0,
\end{align}
where we used $c = R \sinh \alpha_0$.  Multiplying this equation by
$e^{-im\phi} \cos(n\theta)$ (with $n > 0$) and integrating over $\phi$
and $\theta$, we get for any $m \in
\Z$ and $n= 1,2,\ldots$:
\begin{align*}
& A_{m,n}^{+} (1 + 2 \cosh\alpha_0\, p_{m,n}) - A_{m,n-1}^{+} p_{m,n-1}  \\
& - A_{m,n+1}^{+} p_{m,n+1} - \delta_{n,1} A_{m,0}^{+} p_{m,0} = 2R\mu A_{m,n}^{+} .
\end{align*}
In turn, multiplication by $e^{-im\phi}$ and integration over $\phi$
and $\theta$ yield for any $m \in \Z$:
\begin{align*}
& A_{m,0}^{+} (1 + 2 \cosh\alpha_0 p_{m,0}) - A_{m,1}^{+} p_{m,1} = 2R\mu A_{m,0}^{+}.
\end{align*}
Denoting by $\A^{+}_m$ the vector of coefficients $A_{m,n}^{+}$, the
above equations can be written in a matrix form:
\begin{equation}
\A^{+}_m \MM^{+}_m = \mu \A^{+}_m,
\end{equation}
with
\begin{align} \nonumber
& [\MM^{+}_m]_{n',n} = \frac{1}{2R} \biggl[\delta_{n,n'} (1 + 2 \cosh\alpha_0\, p_{m,n}) - \delta_{n,n'+1} p_{m,n'} \\  \label{eq:Mp_ext}
& - \delta_{n,n'-1} p_{m,n'} - \delta_{n,1} \delta_{n',0} p_{m,0} \biggr]  \qquad (n,n' = 0,1,2,\ldots).
\end{align}
Similarly, multiplication of Eq. (\ref{eq:auxil1}) by $e^{-im\phi}
\sin(n\theta)$ and integration over $\theta$ and $\phi$ yield
\begin{align*}
& A_{m,n}^{-} (1 + 2 \cosh\alpha_0 \, p_{m,n}) - A_{m,n-1}^{-} p_{m,n-1} \\
& - A_{m,n+1}^{-} p_{m,n+1} = 2R\mu A_{m,n}^{-} \quad (n = 1,2,\ldots),
\end{align*}
which can also be written in a matrix form
\begin{equation}
\A^{-}_m \MM^{-}_m = \mu \A^{-}_m ,
\end{equation}
with
\begin{align} \nonumber
[\MM^{-}_m]_{n',n} & = \frac{1}{2R} \biggl[\delta_{n,n'} (1 + 2 \cosh\alpha_0\, p_{m,n}) - \delta_{n,n'+1} p_{m,n'} \\  \label{eq:Mm_ext}
& - \delta_{n,n'-1} p_{m,n'} \biggr]  \qquad (n,n' = 1,2,\ldots).
\end{align}
One sees that the systems of linear equations determining the
coefficients $A_{m,n}^{+}$ and $A_{m,n}^{-}$ are decoupled and can
thus be solved separately for each $m$.

Let $\{\mu_{m,0}^{+}, \mu_{m,1}^{+}, \ldots\}$ denote the eigenvalues
of the matrix $\MM^{+}_m$, enumerated in an increasing order by $n =
0,1,\ldots$; the associated {\it left} eigenvectors are denoted as
$\V_{m,n}^{+}$.  Similarly, $\mu_{m,n}^{-}$ and $\V_{m,n}^{-}$ with
$n=1,2,\ldots$ are the $n$-th eigenvalue and the associated {\it left}
eigenvector of the matrix $\MM^{-}_m$.  According to
Eq. (\ref{eq:auxil2}), we get two families of eigenfunctions,
enumerated by the double index $m,n$ and the superscript:
\begin{align}  \nonumber
& V^{+}_{m,n}(\alpha,\theta,\phi) = e^{im\phi} \sqrt{\cosh\alpha - \cos\theta}  \\  \label{eq:Vmnp_torus}
& \qquad \times \sum\limits_{n'=0}^\infty  [\V_{m,n}^{+}]_{n'} \cos(n'\theta) 
\frac{P_{n'-\frac12}^{|m|}(\cosh\alpha)}{P_{n'-\frac12}^{|m|}(\cosh\alpha_0)} 
\end{align}
and
\begin{align}  \nonumber
& V^{-}_{m,n}(\alpha,\theta,\phi) = e^{im\phi} \sqrt{\cosh\alpha - \cos\theta} \\   \label{eq:Vmnm_torus}
& \qquad \times \sum\limits_{n'=1}^\infty  [\V_{m,n}^{-}]_{n'} \sin(n'\theta) 
\frac{P_{n'-\frac12}^{|m|}(\cosh\alpha)}{P_{n'-\frac12}^{|m|}(\cosh\alpha_0)} \,.
\end{align}
The associated eigenvalues are $\mu_{m,n}^{+}$ and $\mu_{m,n}^{-}$,
respectively.  These eigenfunctions respect the reflection symmetry of
the problem (with respect to the horizontal plane):
\begin{equation}
V^{\pm}_{m,n}(\alpha,-\theta,\phi) = \pm V^{\pm}_{m,n}(\alpha,\theta,\phi).
\end{equation}
Moreover, since $p_{-m,n} = p_{m,n}$ according to
Eq. (\ref{eq:pmn_def}), one has $\MM_{-m}^{\pm} = \MM_m^{\pm}$ so that
any eigenvalue $\mu_{m,n}^{\pm}$ with $m \ne 0$ is at least twice
degenerate.  In particular, one can take two linearly independent
combinations of $V^{\pm}_{m,n}$ and $V^{\pm}_{-m,n}$ to form
real-valued eigenfunctions.

For applications, we also need to ensure the normalization of
eigenfunctions and to compute their projection onto a constant.  We
get
\begin{align*}
J_{m,n}^{+} & = \int\limits_{\pa} |V_{m,n}^{+}|^2 = \int\limits_{-\pi}^\pi d\theta \int\limits_0^{2\pi} d\phi \,h_\theta \, h_\phi \,
|V_{m,n}^{+}(\alpha_0,\theta,\phi)|^2 \\
& = 2\pi c^2 \sinh\alpha_0 \sum\limits_{n_1,n_2=0}^\infty [\V_{m,n}^{+}]_{n_1}^* [\V_{m,n}^{+}]_{n_2} \\ 
& \times \int\limits_{-\pi}^\pi d\theta \, \frac{\cos(n_1\theta) \cos(n_2\theta)}{\cosh\alpha_0 - \cos\theta} \,.
\end{align*}
A direct computation of this double sum may be inaccurate and
time-consuming.  To avoid this step, we recall that $V_{m,n}^{+}$
satisfies the Steklov boundary condition and write
\begin{align*}
& J_{m,n}^{+} = \int\limits_{\pa} [V_{m,n}^{+}]^{*} \frac{\partial_n V_{m,n}^{+}}{\mu_{m,n}^{+}}
= \frac{2\pi c \sinh^2\alpha_0}{\mu_{m,n}^{+}} \\
& \times \sum\limits_{n_1,n_2=0}^\infty [\V_{m,n}^{+}]_{n_1}^* [\V_{m,n}^{+}]_{n_2} 
\int\limits_{-\pi}^\pi d\theta \frac{\cos(n_1\theta) \cos(n_2\theta)}{\sqrt{\cosh\alpha_0 - \cos\theta}}  \\
& \times \biggl[\frac{1}{2\sqrt{\cosh\alpha_0 - \cos\theta}} + \sqrt{\cosh \alpha_0 - \cos\theta} \, p_{m,n_2}\biggr],
\end{align*}
where we explicitly evaluated the normal derivative.  In the first
integral, one can recognize the above expression for $J_{m,n}^+$,
whereas the second integral is easily evaluated due to the
orthogonality of cosine functions.  As a consequence, we get
\begin{align*}
J_{m,n}^{+} & = \frac{J_{m,n}^{+}}{2R \mu_{m,n}^{+}} + \frac{2\pi^2 c \sinh^2\alpha_0}{\mu_{m,n}^{+}}
\biggl(2 p_{m,0} \bigl|[\V_{m,n}^{+}]_{0}\bigr|^2 \\
& + \sum\limits_{n'=1}^\infty p_{m,n'} \bigl|[\V_{m,n}^{+}]_{n'}\bigr|^2 \biggr),
\end{align*}
from which $J_{m,n}^{+}$ can be easily expressed:
\begin{align} \label{eq:Jmnp_torus}
J_{m,n}^{+} & = \frac{2\pi^2 c^2 \sinh\alpha_0}{R\mu_{m,n}^{+} - \frac12} \\  \nonumber
& \times \biggl(2 p_{m,0} \bigl|[\V_{m,n}^{+}]_{0}\bigr|^2 + \sum\limits_{n'=1}^\infty p_{m,n'} \bigl|[\V_{m,n}^{+}]_{n'}\bigr|^2 \biggr) ,
\end{align}
where we used $c = R \sinh\alpha_0$.  Similarly, we have
\begin{align} \nonumber
J_{m,n}^{-} & = \int\limits_{\pa} |V_{m,n}^{-}|^2 = \frac{2\pi^2 c^2 \sinh\alpha_0}{R\mu_{m,n}^{-} - \frac12}  
  \sum\limits_{n'=1}^\infty p_{m,n'} \bigl|[\V_{m,n}^{-}]_{n'}\bigr|^2 .
\end{align}

We also evaluate the projection of $V_{m,n}^+$ onto a constant:
\begin{align} \nonumber
I_{m,n}^{+} & = \int\limits_{\pa} V_{m,n}^{+} = \int\limits_{-\pi}^\pi d\theta \int\limits_0^{2\pi} d\phi \,h_\theta \, h_\phi \,
V_{m,n}^{+}(\alpha_0,\theta,\phi) \\   \label{eq:Imnp_torus}
& = 2\pi c^2 \sinh\alpha_0 \delta_{m,0} \sum\limits_{n'=0}^\infty [\V_{m,n}^{+}]_{n'}\, \hat{b}_{n'} ,
\end{align}
where
\begin{equation}  \label{eq:bhat_def}
\hat{b}_n = \int\limits_{-\pi}^{\pi} d\theta \frac{\cos(n\theta)}{(\cosh \alpha_0 - \cos\theta)^{3/2}} \,.
\end{equation}
This integral can be found from the Fourier expansion
\cite{Belevitch83}:
\begin{align} \nonumber
& \sum\limits_{n=0}^\infty \frac{\cos(n\theta)}{1 + \delta_{n,0}} Q_{n-\frac12}^\mu(\cosh\alpha) \\  \label{eq:Qn_Fourier}
& = \frac{e^{i\pi \mu} \sqrt{\pi} \,\Gamma(\mu+\frac12) (\sinh\alpha)^\mu}
{\sqrt{8} \, (\cosh \alpha - \cos\theta)^{\mu+\frac12}}  \,,
\end{align}
which is valid for any $\mu \ne -1/2, -3/2,\ldots$.  Setting $\mu =
1$, one gets
\begin{equation} \label{eq:bhat_Qn}
\hat{b}_n = - 2\sqrt{8} \frac{Q_{n-\frac12}^1(\cosh\alpha_0)}{\sinh\alpha_0} \,.
\end{equation}
In turn, one has $\int\nolimits_{\pa} V_{m,n}^{-} = 0$ due to the
symmetry of the eigenfunction $V_{m,n}^{-}$.  As a consequence, only
the axially symmetric eigenmodes with $m=0$ contribute to the
steady-state concentration and the diffusive flux.  We conclude that
the diffusive flux is determined by the following spectral parameters:
\begin{equation}  \label{eq:Fk_torus}
\mu_k = \mu_{0,k}^+,  \quad F_k = \frac{|I_{0,k}^+|^2}{|\pa|  J_{0,k}^+}  \qquad (k = 0,1,2,\ldots),
\end{equation}
where $|\pa| = 4\pi^2 HR$ is the surface area of the torus, and we
included explicitly the normalization $J_{0,k}^+$.

\subsection{Practical implementation}
\label{sec:torus_practical}

The crucial advantage of the proposed method is the possibility to
construct {\it separately} the eigenvalues and eigenfunctions for each
value of $m$.  For instance, it is sufficient to focus on axially
symmetric eigenmodes ($m = 0$) for the analysis of the diffusive flux
$J$.  In the following, we choose $m$ and then fix a truncation order
$\nmax$.  One needs to construct a truncated matrix $\MM_m^{+}$ of
size $(\nmax+1)\times (\nmax+1)$, and a truncated matrix $\MM_m^{-}$
of size $\nmax \times \nmax$.  These are tridiagonal matrices,
whose elements are expressed via Eqs. (\ref{eq:Mp_ext},
\ref{eq:Mm_ext}) in terms of $p_{m,n}$ given by Eq. (\ref{eq:pmn_def2}).
In turn, the Legendre functions $P_{n-\frac12}^m(z)$ with $z \geq 1$
can be found either from their integral representation \cite{NIST_Pnu}
\begin{equation}
P_\nu^m(z) = \frac{[\nu+1]_m}{\pi} \int\limits_0^\pi d\phi \, \cos(m\phi) \biggl(z + \sqrt{z^2-1}\cos\phi\biggr)^{\nu} ,
\end{equation}
(with $[\nu+1]_m = (\nu+1)(\nu+2)\cdots(\nu+m)$), or from recurrence
relations \cite{Abramowitz}.  Once the truncated matrix $\MM_m^{\pm}$
is constructed, its numerical diagonalization yields an approximation
for a number of eigenvalues $\mu_{m,n}^{\pm}$ and the associated left
eigenvectors $\V_{m,n}^{\pm}$, which determine the Steklov
eigenfunctions via Eqs. (\ref{eq:Vmnp_torus}, \ref{eq:Vmnm_torus}).
Increasing the truncation order $\nmax$, one can ensure the required
accuracy of the computed eigenvalues and eigenfunctions.  Finally, the
coefficients $F_k$ determining the diffusive flux are found according
to Eq. (\ref{eq:Fk_torus}), in which $J_{0,k}^{+}$ and $I_{0,k}^{+}$
are evaluated from the truncated expressions (\ref{eq:Jmnp_torus},
\ref{eq:Imnp_torus}), with $\hat{b}_n$ given by Eq. (\ref{eq:bhat_Qn}).
The latter involves the Legendre function of the second kind, which
can be expressed through the Whipple's formula
\begin{align}  \nonumber
Q_{n-\frac12}^m(z) & = (-1)^m \Gamma(m-n+\tfrac12) \frac{\sqrt{\pi/2}}{(z^2-1)^{\frac14}} \\   \label{eq:Whipple}
& \times P_{m-\frac12}^n\biggl(\frac{z}{\sqrt{z^2-1}}\biggr) .
\end{align}

\begin{figure}
\begin{center}
\includegraphics[width=85mm]{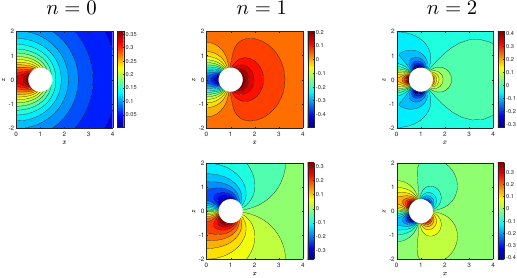}
\end{center}
\caption{
Several axially symmetric Steklov eigenfunctions $V_{0,n}^{\pm}$ for
the torus with $R = 0.5$ and $H = 1$.  The cross-section in the $xz$
plane with $x > 0$, that corresponds to $\phi = 0$, is shown.  Top
row: eigenfunctions $V_{0,n}^{+}$ that are symmetric with respect to
the horizontal plane; bottom row: eigenfunctions $V_{0,n}^{-}$ that
are antisymmetric. }
\label{fig:torus_eigenfunctions}
\end{figure}

Table \ref{tab:muk} presents the first two eigenvalues $\mu_{0,0}^+$
and $\mu_{0,1}^{+}$ as functions of the aspect ratio $R/H$, whereas
Fig. \ref{fig:torus_eigenfunctions} illustrates several axially
symmetric Steklov eigenfunctions for a torus.  The accuracy of these
results was also verified by comparison with another computation based
on a finite-element method described in \cite{Grebenkov25}.

We note that the proposed method is not applicable in the limit $R =
H$ that corresponds to $\alpha_0 = 0$.  According to
Eq. (\ref{eq:pmn_def2}), the coefficient $p_{0,n}$ for any fixed $n$
has a finite limit, $p_{0,n}\to (n^2-1/4)/2$, as $z = H/R \to 1$,
where we used the asymptotic behavior of the Legendre functions:
%
\begin{equation}
P_\nu^m(z) \approx \frac{\Gamma(\nu+m+1)}{m! \, \Gamma(\nu-m+1)} \bigl((z-1)/2\bigr)^{\frac{m}{2}}  \quad (z\to 1^{+}).
\end{equation}
However, as the matrices $\MM_0^{\pm}$ are infinite-dimensional, their
construction requires the elements with large $n$, in which case the
asymptotic behavior is different.  In fact, one can use another
asymptotic relation
%
\begin{equation}
P_\nu^\mu(\cosh \alpha_0) \approx \nu^\mu \sqrt{\frac{\alpha_0}{\sinh\alpha_0}}\, I_\mu((\nu+1/2)\alpha_0)  \quad (\nu\to\infty),
\end{equation}
where $I_\mu(z)$ is the modified Bessel function of the first kind.
As a consequence, one gets in the leading order that $p_{m,n} \sim
(n-\tfrac12)/\sinh \alpha_0$ as $n\to \infty$ for a fixed $\alpha_0$.
One sees that the two limits $R \to H$ and $n\to \infty$ cannot be
inter-changed, i.e., one cannot simply substitute $\alpha_0 = 0$ in
the above relations.  In practice, one can compute the eigenvalues for
any fixed $R < H$, but larger and larger truncation orders may be
needed as $R$ gets closer to $H$.

\subsection{Asymptotic behavior for a thin torus}
\label{sec:torus_thin}

We study the asymptotic behavior of the eigenvalues of the exterior
Steklov problem as $R \to 0$.  In this limit, one has $z =
\cosh\alpha_0 = H/R \to \infty$ and thus $\alpha_0$ goes to infinity.

For the case $n = 0$, it is convenient to use the following
representation \cite{NIST14.5.25}
\begin{equation}  \label{eq:P12_K}
P_{-\frac12}(z) = \frac{2}{\pi} \sqrt{\frac{2}{1+z}}\, K\biggl(\sqrt{\frac{z-1}{z+1}}\biggr),
\end{equation}
where $K(z)$ is the complete elliptic integral of the first kind:
\begin{equation}
K(z) = \int\limits_0^{\pi/2} \frac{d\theta}{\sqrt{1-z^2\sin^2\theta}} \,.
\end{equation}
Substituting the asymptotic behavior of $K(z)$ near $1$
into Eq. (\ref{eq:P12_K}), one gets in the leading order
\begin{equation}
P_{-\frac12}(z) \approx \frac{\sqrt{2} \ln(8z)}{\pi\, \sqrt{z}}   \quad (z\to\infty).
\end{equation}
Since
\begin{equation}
P_\nu^m(z) = (z^2-1)^{\frac{m}{2}} \frac{d^m}{dz^m} P_\nu(z),
\end{equation}
we also get for $m > 0$
\begin{equation}
P_{-\frac12}^m(z) \approx  (-1)^m [1/2]_m \frac{\sqrt{2} \,[\ln(8z) - d_m]}{\pi\, \sqrt{z}}   \quad (z\to\infty),
\end{equation}
where $[a]_m = a(a+1)\cdots (a+n-1)$,
\begin{equation}
d_m = \sum\limits_{j=1}^m \frac{1}{j-1/2}   \qquad (m = 1,2,\ldots),
\end{equation}
and we also set $d_0 = 0$.  Substituting this asymptotic behavior into
Eq. (\ref{eq:pmn_def2}), we find
\begin{equation}
p_{m,0} \approx \frac{1}{z} \biggl(\frac{1}{\ln(8z) - d_m} - \frac12\biggr)  \qquad (z\to\infty).
\end{equation}

In turn, for $n > 0$, one has 
%
\begin{equation}
P_{n-\frac12}^m(z) \simeq \frac{\Gamma(n) (2z)^{n-\frac12}}{\sqrt{\pi}\, \Gamma(n+\frac12-m)}   \quad (z\to \infty),
\end{equation}
so that 
\begin{equation}
p_{m,n} \approx \frac{n-\frac12}{z}  \quad (z\to\infty)
\end{equation}
in the leading order.  

Substituting these asymptotic relations to Eqs. (\ref{eq:Mp_ext},
\ref{eq:Mm_ext}), we see that the leading-order term of the matrix
$R\,\MM^{+}_m$ is the diagonal matrix, whose elements are $1/(\ln (8z)
- d_m), \, 1, \, 2,\, \ldots$, whereas the next-order term is $O(R/H)$
(with eventual logarithmic corrections).  In turn, the leading-order
term of the matrix $R\, \MM^{-}_m$ is the diagonal matrix with the
elements $1,2,\ldots$ on the diagonal.  As a consequence, we obtain
the following asymptotic behavior as $R\to 0$ for any $m\in \Z$:
\begin{align}  \label{eq:mu_ext_torus}
R \mu_{m,0}^{+} & \approx \frac{1}{\ln (H/R) + \ln 8 - d_m} \,, \\
R \mu_{m,n}^{+} & \approx R\mu_{m,n}^{-} \approx n + O(R/H) \quad (n = 1,2,\ldots).
\end{align}
This asymptotic behavior agrees with recent mathematical works, which
studied the Steklov problem for a bounded domain with a thin toroidal
hole \cite{Chiado20} and a tubular excision from a closed manifold
\cite{Brisson22}.

\begin{figure}
\begin{center}
\includegraphics[width=85mm]{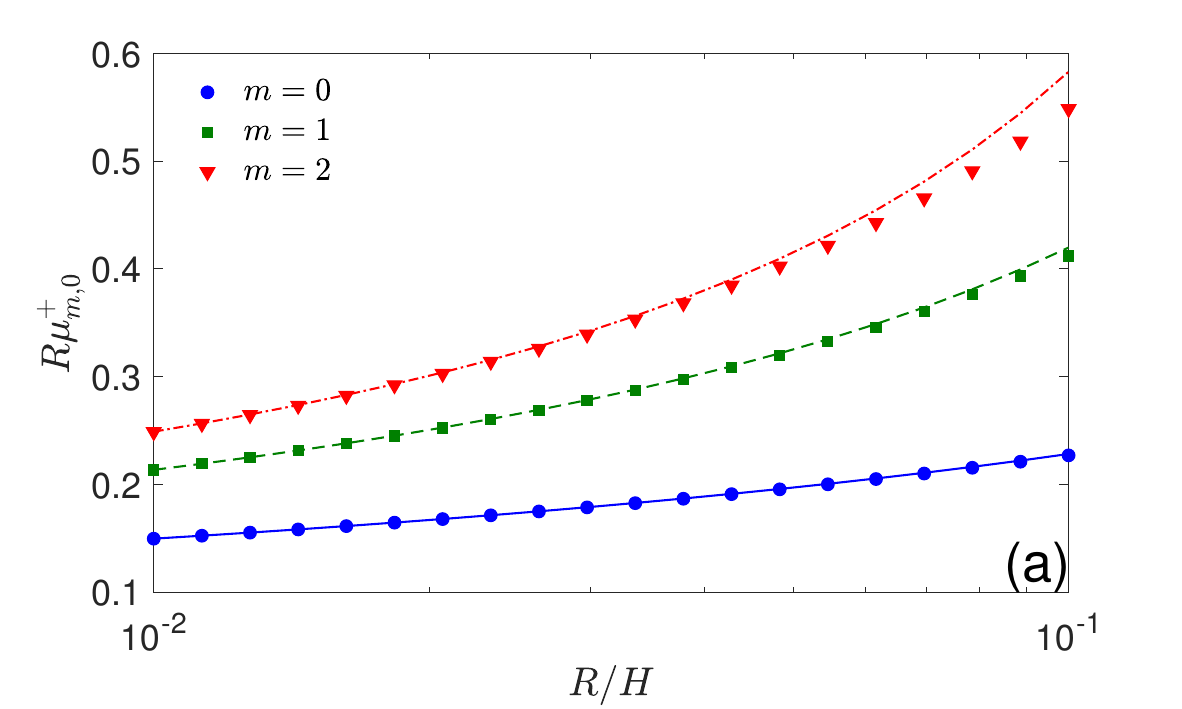} 
\includegraphics[width=85mm]{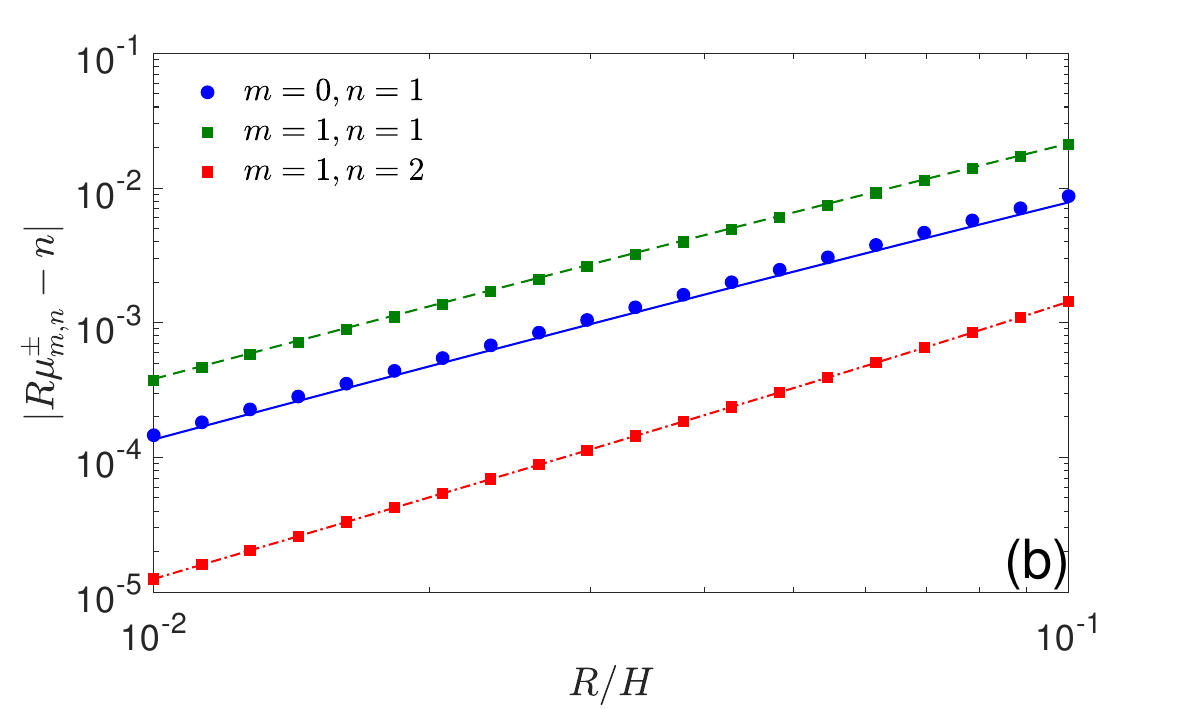} 
\end{center}
\caption{
{\bf (a)} Rescaled eigenvalues $R \mu_{m,0}^+$ of the exterior Steklov
problem for the torus with major and minor radii $H$ and $R$.  Symbols
present the numerical results obtained by diagonalizing the truncated
matrices $\MM^{+}_m$ of size $10\times 10$, whereas lines show the
asymptotic relation (\ref{eq:mu_ext_torus}).  {\bf (b)} The difference
$|R\mu_{m,n}^{\pm} - n|$ between the eigenvalues $\mu_{m,n}^{\pm}$
(rescaled by $R$) and their limits $n$.  Symbols present the results
for $\MM^{+}_m$, lines show the results for $\MM_m^{-}$.  In both
cases, the truncated matrices $\MM^{\pm}_m$ of size $10\times 10$ were
numerically diagonalized. }
\label{fig:mu0_ext_torus}
\end{figure}

Figure \ref{fig:mu0_ext_torus}(a) illustrates the high accuracy of the
asymptotic relation (\ref{eq:mu_ext_torus}) for several $m$.  In
addition, the panel (b) shows that the eigenvalues $\mu_{m,n}^{\pm}$
with $n > 0$, rescaled by $R$, approach their limits $n$ linearly with
$R/H$, in agreement with our predictions.

\subsection{Interior problem}

The computation of the Steklov spectrum for the interior of a torus
(i.e., for $\alpha > \alpha_0$) can be performed in the same way,
except that $P_{n-\frac12}^{|m|}(\cosh\alpha)$ is replaced by
$Q_{n-\frac12}^{|m|}(\cosh\alpha)$, and the direction of the normal
derivative is the opposite.  As a consequence, we obtain two families
of Steklov eigenfunctions
\begin{align}  \nonumber
& V^{+}_{m,n}(\alpha,\theta,\phi) = e^{im\phi} \sqrt{\cosh\alpha - \cos\theta} \\
& \qquad \times \sum\limits_{n'=0}^\infty  [\V_{m,n}^{+}]_{n'} \cos(n'\theta) 
\frac{Q_{n'-\frac12}^{|m|}(\cosh\alpha)}{Q_{n'-\frac12}^{|m|}(\cosh\alpha_0)} 
\end{align}
and
\begin{align}  \nonumber
& V^{-}_{m,n}(\alpha,\theta,\phi) = e^{im\phi} \sqrt{\cosh\alpha - \cos\theta} \\
& \qquad \sum\limits_{n'=1}^\infty  [\V_{m,n}^{-}]_{n'} \sin(n'\theta) 
\frac{Q_{n'-\frac12}^{|m|}(\cosh\alpha)}{Q_{n'-\frac12}^{|m|}(\cosh\alpha_0)} ,
\end{align}
where $\V_{m,n}^{+}$ are the left eigenvectors of the matrix
\begin{align}  \nonumber
& [\MM^{+}_m]_{n',n} = \frac{-1}{2R} \biggl[\delta_{n,n'} (1 + 2 \cosh\alpha_0 q_{m,n}) - \delta_{n,n'+1} q_{m,n'} \\    \label{eq:Mp_int}
& - \delta_{n,n'-1} q_{m,n'} - \delta_{n,1} \delta_{n',0} q_{m,0} \biggr]  \qquad (n,n' = 0,1,2,\ldots),
\end{align}
and $\V_{m,n}^{-}$ are the left eigenvectors of the matrix
\begin{align}  \nonumber
& [\MM^{-}_m]_{n',n} = \frac{-1}{2R} \biggl[\delta_{n,n'} (1 + 2 \cosh\alpha_0 q_{m,n}) - \delta_{n,n'+1} q_{m,n'} \\    \label{eq:Mm_int}
& - \delta_{n,n'-1} q_{m,n'} \biggr]  \qquad (n,n' = 1,2,\ldots),
\end{align}
with
\begin{equation}
q_{m,n} = q_{-m,n} = \frac{Q_{n-\frac12}^{m'}(\cosh\alpha_0)}{Q_{n-\frac12}^{m}(\cosh\alpha_0)}  \quad (m = 0,1,2,\ldots).
\end{equation}
Using the recurrence relation
\begin{equation}
\partial_z Q_\nu^\mu(z) = \frac{1}{z^2-1} \biggl(\sqrt{z^2-1} Q_\nu^{\mu+1}(z) + \mu z Q_\nu^\mu(z)\biggr),
\end{equation}
one gets
\begin{equation}
q_{m,n} = \frac{1}{\sinh\alpha_0} \left(\frac{Q_{n-\frac12}^{m+1}(\cosh\alpha_0)}{Q_{n-\frac12}^m(\cosh\alpha_0)} + m\, \ctanh\alpha_0 \right).
\end{equation}
The involved Legendre functions of the second kind can be found via
the Whipple's formula (\ref{eq:Whipple}), integral representations, or
recurrence relations \cite{Abramowitz}.  Despite the general interest
in the spectral properties of the Steklov problem for the interior of
a torus, further development of this topic is beyond the scope of the
paper.

\subsection{Perfectly reactive torus}
\label{sec:torus_Rc}

For completeness, we reproduce the solution for a perfectly reactive
torus (see more details in \cite{Belevitch83}).  The axial symmetry of
this problem implies $C_A(\x) = C_A(1 - u(\x))$, where the harmonic
function $u(\x)$ can be searched in the form
\begin{equation}
u = \sqrt{\cosh\alpha-\cos\theta} \sum\limits_{n=0}^\infty b_n \cos(n\theta) \frac{P_{n-\frac12}(\cosh\alpha)}{P_{n-\frac12}(\cosh\alpha_0)}  \,.
\end{equation}
The coefficients $b_n$ are fixed by the Dirichlet boundary
condition $u|_{\pa} = 1$:
\begin{equation}  \label{eq:bn}
b_n = \frac{\epsilon_n}{\pi} \int\limits_{-\pi}^\pi d\theta \, \frac{\cos(n\theta)}{\sqrt{\cosh\alpha_0 - \cos\theta}} 
= \frac{\epsilon_n \sqrt{8}}{\pi} Q_{n-\frac12}(\cosh\alpha_0) ,
\end{equation}
where the second equality was obtained from the Fourier expansion
(\ref{eq:Qn_Fourier}) by setting $\mu = 0$, and $\epsilon_n =
1/(1+\delta_{n,0})$.  As a consequence, the diffusive flux is
\begin{align*}
J &= DC_A \int\limits_{\pa} \partial_n u = D C_A 2\pi \int\limits_{-\pi}^\pi d\theta h_\theta h_\phi \frac{1}{h_\alpha} \partial_\alpha u \\
& = D C_A 2\pi c \sinh\alpha_0^2 \int\limits_{-\pi}^\pi d\theta \frac{1}{\cosh\alpha_0 - \cos\theta} \sum\limits_{n=0}^\infty b_n \cos(n\theta) \\
& \times \biggl[\frac{1}{2\sqrt{\cosh\alpha_0-\cos\theta}} + \sqrt{\cosh\alpha_0 - \cos\theta}\, p_{0,n} \biggr]  \\
& = D C_A 2\pi c \sinh\alpha_0^2 \sum\limits_{n=0}^\infty b_n  \biggl[\frac12 \hat{b}_n + \frac{\pi}{\epsilon_n} b_n p_{0,n}\biggr],
\end{align*}
where $\hat{b}_n$ was defined by Eq. (\ref{eq:bhat_def}).
Substituting Eqs. (\ref{eq:pmn_def}, \ref{eq:bhat_Qn}, \ref{eq:bn})
and using the Wronskian of Legendre functions,
%
\begin{equation}
P_\nu^\mu(z) \partial_z Q_\nu^\mu(z) - Q_\nu^\mu(z) \partial_z P_\nu^\mu(z) = \frac{-e^{i\mu \pi} \Gamma(\nu+\mu+1)}{\Gamma(\nu-\mu+1) (z^2-1)} \,,
\end{equation}
one retrieves Eq. (\ref{eq:Rc_torus}) for the capacitance of a torus.

\section{Pair of spheres}
\label{sec:bisphere}

In this Appendix, we replicate the computation of Appendix
\ref{sec:torus} to a pair of spheres.  For this purpose, we employ
bispherical coordinates \cite{Morse} 
%
\begin{equation}
\left(\begin{array}{c} x \\ y \\ z \\ \end{array}\right) =
\frac{c}{\cosh \alpha - \cos\theta}  \left(\begin{array}{c} \sin \theta \cos\phi \\ \sin\theta \sin\phi \\ \sinh\alpha \\ \end{array}\right),
\end{equation}
where $-\infty < \alpha < +\infty$, $0\leq \theta \leq \pi$, $0 \leq
\phi < 2\pi$, and $c > 0$.  They can be expressed in terms of
Cartesian coordinates as
\begin{equation}
\alpha = \sinh^{-1} \biggl(\frac{2cz}{Q}\biggr),  \qquad
\theta = \cos^{-1} \biggl(\frac{P^2 - c^2}{Q}\biggr), 
\end{equation}
and $\phi = \tan^{-1}(y/x)$, with
\begin{equation}
P = \sqrt{x^2+y^2+z^2}, \quad Q = \sqrt{(P^2+c^2)^2 - (2cz)^2} \,,
\end{equation}
whereas the scale factors are
\begin{equation}
h_\alpha = h_\theta = \frac{c}{\cosh\alpha - \cos\theta}, \quad h_\phi = \frac{c\, \sin\theta}{\cosh\alpha - \cos\theta} \,.
\end{equation}

For a given $\alpha_0 > 0$, the domain $\Omega = \{ \alpha >
\alpha_0\}$ is the ball of radius $R = c/\sinh(\alpha_0)$, centered at
$(0,0, H)$, with $H = c\, \ctanh(\alpha_0) > 0$.  Similarly, for
$\alpha_0 < 0$, the domain $\Omega = \{ \alpha < \alpha_0\}$ is the
ball of radius $c/|\sinh(\alpha_0)|$, centered at $(0,0, -H)$, with $H
= c\, |\ctanh(\alpha_0)|$.  In turn, the exterior of these two equal
balls is given by $|\alpha | < \alpha_0$.  In the next subsection, we
focus on this setting of balls of equal radii, for which computations
are simpler.  The exterior of two balls of different radii will be
discussed in Sec. \ref{sec:two_extension}.

\subsection{Exterior Steklov problem}

In this subsection, we consider the exterior of two equal balls of
radii $R$, whose centers are located at $(0,0,\pm H)$.  The boundary
$\pa$ of the domain is composed of two spheres, denoted as
$\pa_{\pm}$.  In the bispherical coordinates, this domain is defined
as $|\alpha| < \alpha_0$, where
\begin{align*}
c & = R \sinh \alpha_0, \quad \cosh \alpha_0 = \frac{H}{R} > 1, \\
\alpha_0 & = \ln\bigl(H/R + \sqrt{(H/R)^2 - 1}\bigr) > 0,
\end{align*}
so that $H = c\, \ctanh(\alpha_0)$ and $c = \sqrt{H^2-R^2}$.

Due to a partial separation of variables in the bispherical
coordinates, we search a solution of Eq. (\ref{eq:Steklov}) in the
form:
\begin{align}  \nonumber  
V & = \sqrt{\cosh\alpha - \cos\theta}  \sum\limits_{n=0}^\infty \sum\limits_{m=-n}^n 
\biggl[A_{m,n}^{-} \frac{\sinh((n+\frac12)\alpha)}{\sinh((n+\frac12)\alpha_0)} \\   \label{eq:V_twospheres}
& +  A_{m,n}^{+} \frac{\cosh((n+\frac12)\alpha)}{\cosh((n+\frac12)\alpha_0)}\biggr] Y_{m,n}(\theta,\phi),
\end{align}
where $Y_{m,n}(\theta,\phi)$ are normalized spherical harmonics:
\begin{subequations}
\begin{align}  \label{eq:Ymn_def}
Y_{m,n}(\theta,\phi) &= a_{m,n} P_n^m(\cos\theta) e^{im\phi} , \\
a_{m,n} & = \sqrt{\frac{2n+1}{4\pi} \, \frac{(n-m)!}{(n+m)!}} \,,
\end{align}
\end{subequations}
$P_n^m(z)$ are the associated Legendre polynomials, and
$A_{m,n}^{\pm}$ are unknown coefficients (we adopt the convention that
$a_{m,n} \equiv 0$ for any $|m| > n$).

We first impose the Steklov boundary condition on the sphere $\pa_+$:
\begin{align*}
& \partial_n V|_{\pa_+} = \frac{1}{h_{\alpha_0}} (\partial_\alpha V)|_{\alpha = \alpha_0} 
= \frac{\sqrt{\cosh\alpha_0 - \cos\theta}}{2c} \\
& \times \biggl(\sinh\alpha_0 \sum\limits_{m,n'} \bigl[A_{m,n'}^{-}  
+ A_{m,n'}^{+} \bigr] Y_{m,n'}(\theta,\phi) \\
& + (\cosh\alpha_0 - \cos\theta) \sum\limits_{m,n'} \bigl[A_{m,n'}^{-} C^{-}_{n'} 
 + A_{m,n'}^{+} C^{+}_{n'}\bigr] Y_{m,n'}\biggr) \\
& = \mu V|_{\pa_+} = \mu \sqrt{\cosh\alpha_0 - \cos\theta}  \sum\limits_{m,n'} \bigl[A_{m,n'}^{-} + 
A_{m,n'}^{+} \bigr] Y_{m,n'},
\end{align*}
where we used the shortcut notation $\sum\nolimits_{m,n'}$ for the sum
over $n = 0,1,2,\ldots$ and $m = -n,-n+1,\ldots,n$, and
\begin{subequations}
\begin{align}
C^{-}_n & = (2n+1)\, \ctanh((n+\tfrac12)\alpha_0),  \\ 
C^{+}_n & = (2n+1) \tanh((n+\tfrac12)\alpha_0).
\end{align}
\end{subequations}
Multiplication of the above relation by $Y_{m,n}^*(\theta,\phi)
\sin\theta$ and integration over $\theta$ and $\phi$ yield
\begin{align} \nonumber
& \sinh\alpha_0 \bigl[A_{m,n}^{-} +  A_{m,n}^{+}\bigr] + \cosh\alpha_0 \bigl[A_{m,n}^{-} C^{-}_n + A_{m,n}^{+} C^{+}_n\bigr] \\  \nonumber
&  - \sum\limits_{n'=0}^\infty \biggl(A_{m,n'}^{-} C^{-}_{n'} 
+ A_{m,n'}^{+} C^{+}_{n'}\biggr) [\mm_m]_{n,n'} \\   \label{eq:cond_pap}
&  = 2c \mu \bigl[A_{m,n}^{-}  + A_{m,n}^{+} \bigr] ,
\end{align}
where
\begin{align} \nonumber
[\mm_m]_{n,n'} & = 2\pi a_{m,n} a_{m,n'} \int\limits_{-1}^1 dx \, x \, P_n^m(x) \, P_{n'}^m(x).
\end{align}
Using the Gaunt's formula, one can evaluate the integral, which takes
nonzero values only at $|n-n'| = 1$:
\begin{equation}
\int\limits_{-1}^1 dx \, x \, P_{n-1}^m(x) \, P_n^m(x) = (-1)^m \frac{2(n-m) \, (n+m)!}{(4n^2-1) \, (n-m)!} \,,
\end{equation}
from which
\begin{equation}
[\mm_m]_{n,n'} = \delta_{n,n'+1} \eta_{m,n} + \delta_{n,n'-1} \eta_{m,n'}   \quad (n,n' \geq |m|),
\end{equation}
where
\begin{equation}  
\eta_{m,n} = (-1)^m \sqrt{\frac{n^2-m^2}{4n^2-1}} \,.
\end{equation}

At the boundary $\pa_-$, which corresponds to $\alpha = -\alpha_0$, we
get a similar boundary condition with the opposite sign
\begin{align*}
& \partial_n V|_{\pa_-} = - \frac{1}{h_{\alpha_0}} (\partial_\alpha V)|_{\alpha = -\alpha_0} 
= - \frac{\sqrt{\cosh\alpha_0 - \cos\theta}}{2c} \\
& \times \biggl(\sinh\alpha_0 \sum\limits_{m,n'} \bigl[A_{m,n'}^{-}  
- A_{m,n'}^{+} \bigr] Y_{m,n'}(\theta,\phi) \\
& + (\cosh\alpha_0 - \cos\theta) \sum\limits_{m,n'} \bigl[A_{m,n'}^{-} C^{-}_{n'} 
 - A_{m,n'}^{+} C^{+}_{n'} \bigr] Y_{m,n'}\biggr) \\
& = \mu \sqrt{\cosh\alpha_0 - \cos\theta}  \sum\limits_{m,n'} \bigl[- A_{m,n'}^{-} + 
A_{m,n'}^{+} \bigr] Y_{m,n'},
\end{align*}
from which we obtain another set of linear equations:
\begin{align}  \nonumber
& \sinh\alpha_0 \bigl[A_{m,n}^{-} - A_{m,n}^{+}\bigr] + \cosh\alpha_0 \bigl[A_{m,n}^{-} C^{-}_n - A_{m,n}^{+} C^{+}_n\bigr] \\   \nonumber
& - \sum\limits_{n'=0}^\infty \bigl[A_{m,n'}^{-} C^{-}_{n'}  
 - A_{m,n'}^{+} C^{+}_{n'}\bigr] [\mm_m]_{n,n'} \\     \label{eq:cond_pap2}
&  = 2c \mu \bigl[A_{m,n}^{-} - A_{m,n}^{+} \bigr] .
\end{align}
Taking the sum and difference of Eqs. (\ref{eq:cond_pap},
\ref{eq:cond_pap2}), we obtain two decoupled systems:
\begin{align*}
& A_{m,n}^{-} \bigl[\sinh\alpha_0 + \cosh\alpha_0 C^{-}_n\bigr] \\ 
& - \sum\limits_{n'=0}^\infty A_{m,n'}^{-}  C^{-}_{n'} [\mm_m]_{n',n} 
= 2c \mu A_{m,n}^{-} 
\end{align*}
and
\begin{align*}
& A_{m,n}^{+} \bigl[\sinh\alpha_0 + \cosh\alpha_0 C^{+}_n \bigr] \\ 
& - \sum\limits_{n'=0}^\infty A_{m,n'}^{+} C^{+}_{n'} [\mm_m]_{n',n} = 2c \mu A_{m,n}^{+} .
\end{align*}
Denoting by $\A^{\pm}_m$ the vector of coefficients $A_{m,n}^{\pm}$,
we represent the above systems in a matrix form as
\begin{equation}
\A^{+}_m \MM^+_m = \mu \A^{+}_m, \qquad \A^{-}_m \MM^-_m = \mu \A^{-}_m, 
\end{equation}
where
\begin{align} \nonumber
[\MM^{\pm}_m]_{n',n} & = \frac{1}{2c} \biggl(\delta_{n,n'} \bigl(\sinh\alpha_0 + \cosh\alpha_0 C^{\pm}_n\bigr) \\
& - C^{\pm}_{n'} [\mm_m]_{n',n}\biggr)  \qquad (n,n' \geq |m|).
\end{align}
As a consequence, the eigenpairs of the exterior Steklov problem can
be determined separately for different $m$ by diagonalizing the
matrices $\MM^{\pm}_m$.  Denoting by $\mu_{m,n}^{\pm}$ and
$\V_{m,n}^{\pm}$ the eigenvalues and {\it left} eigenvectors of the
matrix $\MM^{\pm}_m$ (enumerated by $n = 0,1,2,\ldots$), we realize
that $\mu_{m,n}^{\pm}$ are the Steklov eigenvalues, whereas the
associated eigenfunctions are given by
\begin{align} \nonumber
& V_{m,n}^{+}(\alpha,\theta,\phi) = \sqrt{\cosh\alpha - \cos\theta}  \\
& \times \sum\limits_{n'=|m|}^\infty [\V_{m,n}^{+}]_{n'} \frac{\cosh((n'+\frac12)\alpha)}{\cosh((n'+\frac12)\alpha_0)} Y_{m,n'}(\theta,\phi)
\end{align}
and 
\begin{align} \nonumber
& V_{m,n}^{-}(\alpha,\theta,\phi) = \sqrt{\cosh\alpha - \cos\theta}   \\
& \times \sum\limits_{n'=|m|}^\infty [\V_{m,n}^{-}]_{n'} \frac{\sinh((n'+\frac12)\alpha)}{\sinh((n'+\frac12)\alpha_0)} Y_{m,n'}(\theta,\phi).
\end{align}
These eigenfunctions respect the reflection symmetry of the problem
(with respect to the horizontal plane):
\begin{equation}
V_{m,n}^{\pm}(-\alpha,\theta,\phi) = \pm V_{m,n}^{\pm}(\alpha,\theta,\phi).
\end{equation}

For applications, we also need to ensure the normalization of
eigenfunctions and to compute their projection onto a constant.  As
the boundary $\pa$ has two components $\pa_{\pm}$ corresponding to
$\alpha_0$ and $-\alpha_0$, we get
\begin{align*}
J_{m,n}^{\pm} & = \int\limits_{\pa} |V_{m,n}^{\pm}|^2 = \int\limits_0^\pi d\theta \int\limits_0^{2\pi} d\phi \,h_\theta \, h_\phi \,
\biggl(|V_{m,n}^{\pm}(\alpha_0,\theta,\phi)|^2 \\
& + |V_{m,n}^{\pm}(-\alpha_0,\theta,\phi)|^2\biggr) \\
& = 4\pi c^2 \sum\limits_{n_1,n_2=0}^\infty [\V_{m,n}^{\pm}]_{n_1}^* [\V_{m,n}^{\pm}]_{n_2} \\
& \times a_{m,n} a_{m,n'} \int\limits_{-1}^1 \frac{dx}{\cosh\alpha_0 - x} \, P_{n_1}^m(x) \, P_{n_2}^m(x).
\end{align*}
To avoid a numerical computation of this integral and the double sum,
we employ the Steklov condition to represent $J_{m,n}^{+}$ as
\begin{widetext}
\begin{align*}
J_{m,n}^{+} & = \int\limits_{\pa} [V_{m,n}^{+}]^* \,  \frac{\partial_n V_{m,n}^{+}}{\mu_{m,n}^{+}} 
 = \frac{2}{\mu_{m,n}^{+}}\int\limits_0^\pi d\theta \int\limits_0^{2\pi} d\phi  \, h_\phi \, 
\biggl([V_{m,n}^{+}]^* \, \partial_\alpha V_{m,n}^{+}\biggr)\biggr|_{\alpha=\alpha_0} 
 = \frac{2\pi c}{\mu_{m,n}^{+}} \sum\limits_{n_1,n_2=0}^\infty [\V_{m,n}^{+}]_{n_1}^* [\V_{m,n}^{+}]_{n_2} a_{m,n_1} a_{m,n_2} \\
& \times \int\limits_{-1}^1 \frac{dx \, P_{n_1}^m(x)\, P_{n_2}^m(x)}{\sqrt{\cosh\alpha_0 - x}} 
\biggl[\frac{\sinh\alpha_0}{\sqrt{\cosh\alpha_0-x}} + \sqrt{\cosh\alpha_0-x} \, C_{n_2}^{+} \biggr]
 =\frac{\sinh\alpha_0}{2c \mu_{m,n}^{+}} J_{m,n}^{+} + \frac{c}{\mu_{m,n}^{+}} 
\sum\limits_{n'=0}^\infty \bigl|[\V_{m,n}^{+}]_{n'}\bigr|^2  C^{+}_{n'},
\end{align*}
\end{widetext}
where we used the orthogonality of the associated Legendre
polynomials.  As a consequence, we get
\begin{align}
J_{m,n}^{+} & = \frac{c}{\mu_{m,n}^{+} - \frac{1}{2R}} \sum\limits_{n'=0}^\infty 
\bigl|[\V_{m,n}^{+}]_{n'}\bigr|^2  C^{+}_{n'}.
\end{align}
Similarly, we have
\begin{align}
J_{m,n}^{-} & = \frac{c}{\mu_{m,n}^{-} - \frac{1}{2R}} \sum\limits_{n'=0}^\infty 
\bigl|[\V_{m,n}^{-}]_{n'}\bigr|^2  C^{-}_{n'} .
\end{align}

In addition, we evaluate
\begin{align} \nonumber
I_{m,n}^{+} & = \int\limits_{\pa} V_{m,n}^{+} = \int\limits_0^\pi d\theta \int\limits_0^{2\pi} d\phi \,h_\theta \, h_\phi \,
\biggl(V_{m,n}^{+}(\alpha_0,\theta,\phi) \\  \nonumber
& + V_{m,n}^{+}(-\alpha_0,\theta,\phi)\biggr) \\
& = 2c^2 \sqrt{2\pi}\, \delta_{m,0} \sum\limits_{n'=0}^\infty [\V_{0,n}^{+}]_{n'} g_{n'}(\cosh \alpha_0), 
\end{align}
where
\begin{equation}
g_n(z) = \sqrt{n+1/2} \int\limits_{-1}^1 \frac{dx}{(z - x)^{3/2}} \, P_{n}(x),
\end{equation}
whereas $I_{m,n}^{-} = 0$ due to the symmetry of the eigenfunctions
$V_{m,n}^{-}$.  Taking the derivative with respect to $\alpha$ of the
following expansion,
\begin{equation}  \label{eq:twospheres_expansion}
\frac{1}{\sqrt{\cosh \alpha - \cos\theta}} = \sqrt{2} \sum\limits_{n=0}^\infty e^{-(n+\frac12)\alpha} P_n(\cos\theta),
\end{equation}
we obtain the expression for the integral:
\begin{equation}
g_n(\cosh\alpha_0) = 2\sqrt{2n+1} \, \frac{e^{-(n+\frac12)\alpha_0}}{\sinh\alpha_0}\, .
\end{equation}

We conclude that the diffusive flux is determined by
\begin{equation}
\mu_k = \mu_{0,k}^{+}, \quad F_k = \frac{|I_{0,k}^{+}|^2}{|\pa| J_{0,k}^{+}}  \quad (k = 0,1,2,\ldots),
\end{equation}
where $|\pa| = 8 \pi R^2$.  The first values of $\mu_k$ and $F_k$ are
reported in Table \ref{tab:muk}.
A practical implementation follows the same lines as in
Sec. \ref{sec:torus_practical}.

\subsection{Perfectly reactive target}

For completeness, we reproduce the steady-state concentration and the
diffusive flux onto a pair of perfectly reactive spheres from
\cite{Samson77}.  Using the expansion (\ref{eq:twospheres_expansion}),
one easily gets:
\begin{align} \nonumber
C_A(\x) & = C_A \biggl(1 - \sqrt{2(\cosh \alpha - \cos\theta)} \\
& \times  \sum\limits_{n=0}^\infty e^{-(n+\frac12)\alpha_0} 
\frac{\cosh((n+\frac12)\alpha)}{\cosh((n+\frac12)\alpha_0)} P_n(\cos\theta) \biggr).
\end{align}
As a consequence, the diffusive flux reads
\begin{align*}
J & = 2DC_A \, 2\pi c \sqrt{2} \sum\limits_{n=0}^\infty e^{-(n+1/2)\alpha_0} 
\int\limits_{-1}^1 \frac{dx\, P_n(x)}{\cosh\alpha_0 - x} \\
& \times \biggl(\frac{\sinh\alpha_0}{2\sqrt{\cosh\alpha_0-x}}
+ \frac12 \sqrt{\cosh\alpha_0-x} \, C_n^{+}\biggr) \\
& = 8\pi DC_A c \sum\limits_{n=0}^\infty e^{-(2n+1)\alpha_0} \bigl(1 + \tanh((n+\tfrac12)\alpha_0)\bigr)  \\
& = 16\pi DC_A c \sum\limits_{n=0}^\infty \frac{e^{-(2n+1)\alpha_0}}{1 + e^{-(2n+1)\alpha_0}} \,,
\end{align*}
from which the capacitance follows as
\begin{equation}
R_c = 4R\sinh\alpha_0 \sum\limits_{n=0}^\infty \frac{e^{-(2n+1)\alpha_0}}{1 + e^{-(2n+1)\alpha_0}}  \,.
\end{equation}
Using a geometric series representation, one recovers the expression
(\ref{eq:Samson}) by Samson and Deutch \cite{Samson77}.  It is worth
noting that this expression differs from the conventional form of the
capacitance $C$ of a pair of equal spheres that is found in textbooks
on electrostatics:
%
\begin{equation}
C = \frac{R}{2} \sum\limits_{n=1}^\infty \frac{\sinh(\alpha_0)}{\sinh(n\alpha_0)}  \,.
\end{equation}
In fact, the latter expression represents the capacitance of a
capacitor when the difference of voltage is applied to two spheres.
In turn, the expression (\ref{eq:Samson}) corresponds to the voltage
applied between the grounded pair of spheres and infinity.

\subsection{Extension to spheres of different radii}
\label{sec:two_extension}

The above computation can be extended to the general case of the
exterior of two balls of radii $R_1$ and $R_2$.  Here we denote two
components of the boundary as $\pa_1$ and $\pa_2$ that are defined by
fixing $\alpha_1$ and $\alpha_2$, respectively.  For convenience, we
assume that $0 < |\alpha_1| < \alpha_2$, i.e., $R_2 < R_1$ and the
second sphere $\pa_2$ lies above the horizontal plane.  One can
distinguish two settings:

(i) If $\alpha_1$ is positive, one deals with a bounded domain
$\Omega$ between two non-concentric spheres $\Omega_2$ and $\Omega_1$
of radii $R_i = c/\sinh(\alpha_i)$, centered at $(0,0,H_i)$, with $H_i
= c \, \ctanh(\alpha_i) > 0$.  The parameters $c$, $\alpha_1$ and
$\alpha_2$ can be determined by fixing the radii $R_1$ and $R_2$ and
the distance between the centers $H = H_1 - H_2 > 0$.  Setting
\begin{equation}
h_1 = \frac12 \biggl(\frac{R_1^2-R_2^2}{H^2} + 1\biggr),  \qquad h_2 = \frac12 \biggl(\frac{R_1^2-R_2^2}{H^2} - 1\biggr),
\end{equation}
such that $H_i = h_i H$, we get
\begin{align*}
x_i & = h_i \frac{H}{R_i} , \quad \alpha_i = \ln\biggl(x_i + \sqrt{x_i^2-1}\biggr),  \\
c & = R_1 \sinh \alpha_1 = R_2 \sinh \alpha_2.
\end{align*}

(ii) In turn, if $\alpha_1 < 0$, the sphere of radius $R_1$ lies below
the horizontal plane, with $H_1 = c \, \ctanh(\alpha_1) < 0$, and one
deals with an unbounded domain $\Omega$.  The parameters $c$,
$\alpha_1$ and $\alpha_2$ can be determined by fixing the radii $R_1$
and $R_2$ and the distance between the centers $H = |H_1| + H_2$.
Setting
\begin{equation}
h_1 = \frac12 \biggl(1 + \frac{R_1^2-R_2^2}{H^2}\biggr),  \qquad h_2 = \frac12 \biggl(1 - \frac{R_1^2-R_2^2}{H^2}\biggr),
\end{equation}
such that $|H_i| = h_i H$, we get
\begin{align*}
x_i & = h_i \frac{H}{R_i} , \quad \alpha_1 = - \ln\biggl(x_1 + \sqrt{x_1^2-1}\biggr),  \\
\alpha_2 & = \ln\biggl(x_2 + \sqrt{x_2^2-1}\biggr), \quad c = R_1 \sinh|\alpha_1| = R_2 \sinh \alpha_2.
\end{align*}

As previously, we search a solution of Eq. (\ref{eq:Steklov}) in the
form (\ref{eq:V_twospheres}), in which $\alpha_0$ is replaced by
$\alpha_2$.  The boundary condition on $\pa_2$ remains unchanged and
yields the system (\ref{eq:cond_pap}) that we rewrite as
\begin{align*} 
& A_{m,n}^{-} \bigl(\sinh\alpha_2 + \cosh\alpha_2 C_n^{-} \bigr) 
 + A_{m,n}^{+} \bigr(\sinh\alpha_2 + \cosh\alpha_2 C_n^{+}\bigr) \\  
& - \sum\limits_{n'=0}^\infty \bigl[A_{m,n'}^{-} C_{n'}^{-} 
 + A_{m,n'}^{+} C_{n'}^{+}\bigr] [\mm_m]_{n',n}
 = 2c \mu \bigl[A_{m,n}^{-}  + A_{m,n}^{+} \bigr] ,
\end{align*}
where
\begin{subequations}
\begin{align}
C_n^{+} & = (2n+1) \tanh((n+\tfrac12)\alpha_2), \\
C_n^{-} & = (2n+1)\, \ctanh((n+\tfrac12)\alpha_2).
\end{align}
\end{subequations}

In turn, the Steklov condition on $\pa_1$ becomes
\begin{align*}
& \partial_n V|_{\pa_1} = - \frac{1}{h_{\alpha_1}} (\partial_\alpha V)_{|\alpha = \alpha_1} 
= - \frac{\sqrt{\cosh\alpha_1 - \cos\theta}}{2c} \\
& \times \biggl[\sinh\alpha_1 \sum\limits_{m,n'} 
\bigl[A_{m,n'}^{-} C_{n'}^{ss} + A_{m,n'}^{+} C_{n'}^{cc} \bigr] Y_{m,n'}(\theta,\phi) \\
& + (\cosh\alpha_1 - \cos\theta) \sum\limits_{m,n'}  
\bigl[A_{m,n'}^{-} C_{n'}^{cs} + A_{m,n'}^{+} C_{n'}^{sc} \bigr] Y_{m,n'}\biggr] \\
& 
= \mu \sqrt{\cosh\alpha_1 - \cos\theta}  \sum\limits_{m,n'} 
\bigl[A_{m,n'}^{-} C_{n'}^{ss} + A_{m,n'}^{+} C_{n'}^{cc} \bigr] Y_{m,n'} ,
\end{align*}
where
\begin{subequations}
\begin{align}
C_n^{ss} & = \frac{\sinh((n+\frac12)\alpha_1)}{\sinh((n+\frac12)\alpha_2)} \,, \quad
C_n^{cc} = \frac{\cosh((n+\frac12)\alpha_1)}{\cosh((n+\frac12)\alpha_2)} \,, \\
C_n^{sc} & = (2n+1)\frac{\sinh((n+\frac12)\alpha_1)}{\cosh((n+\frac12)\alpha_2)} \,, \\
C_n^{cs} & = (2n+1)\frac{\cosh((n+\frac12)\alpha_1)}{\sinh((n+\frac12)\alpha_2)} \,.
\end{align}
\end{subequations}
As a consequence, we get
\begin{align*}
& A_{m,n}^{-} \bigl(\sinh\alpha_1 C_{n}^{ss} + \cosh\alpha_1 C_{n}^{cs}\bigr) \\
& + A_{m,n}^{+} \bigl(\sinh\alpha_1 C_{n}^{cc} + \cosh\alpha_1 C_{n}^{sc}\bigr) \\
& - \sum\limits_{n'=0}^\infty \bigl[A_{m,n'}^{-} C_{n'}^{cs} + A_{m,n'}^{+} C_{n'}^{sc} \bigr] [\mm_m]_{n',n} \\ 
&  = -2c\mu \bigl[A_{m,n}^{-} C_{n}^{ss} + A_{m,n}^{+} C_{n}^{cc} \bigr] .
\end{align*}
In this setting, two systems of linear equations are not decoupled but
can still be written in a matrix form using a $2\times2$ blocked
structure:
\begin{equation}
\left(\begin{array}{c c} \A^-_m & \A^+_m \\ \end{array}\right) 
\underbrace{\left(\begin{array}{c c} \MM^{--}_m & \MM^{-+}_m \\ \MM^{+-}_m & \MM^{++}_m \\ \end{array}\right)}_{=\MM_m}
= \mu \left(\begin{array}{c} \A^-_m \\ \A^+_m \\ \end{array}\right) ,
\end{equation}
where
\begin{widetext}
\begin{align*}
[\MM^{--}_m]_{n',n} & = \frac{1}{2c(C_n^{cc} - C_n^{ss})} \biggl[\delta_{n,n'} \biggl(C_n^{cc} \bigl[\sinh\alpha_2 
+ \cosh\alpha_2 C_n^{-}\bigr] + C_n^{ss} \sinh\alpha_1 + \cosh\alpha_1 C_n^{cs}\biggr)  
 - \bigl[C_n^{cc} C_{n'}^{-} + C_{n'}^{cs} \bigr] [\mm_m]_{n',n}  \biggr], \\
[\MM^{+-}_m]_{n',n} & = \frac{1}{2c(C_n^{cc} - C_n^{ss})} \biggl[\delta_{n,n'} \biggl(C_n^{cc} \bigl[\sinh\alpha_2 
+ \cosh\alpha_2 C_n^{+}\bigr] + C_n^{cc} \sinh\alpha_1 + \cosh\alpha_1 C_n^{sc}\biggr)  
 - \bigl[C_n^{cc} C_{n'}^{+} + C_{n'}^{sc} \bigr] [\mm_m]_{n',n}  \biggr], \\
[\MM^{-+}_m]_{n',n} & = \frac{1}{2c(C_n^{ss} - C_n^{cc})} \biggl[\delta_{n,n'} \biggl(C_n^{ss} \bigl[\sinh\alpha_2 
+ \cosh\alpha_2 C_n^{-}\bigr] + C_n^{ss} \sinh\alpha_1 + \cosh\alpha_1 C_n^{cs}\biggr)  
 - \bigl[C_n^{ss} C_{n'}^{-} + C_{n'}^{cs} \bigr] [\mm_m]_{n',n}  \biggr], \\
[\MM^{++}_m]_{n',n} & = \frac{1}{2c(C_n^{ss} - C_n^{cc})} \biggl[\delta_{n,n'} \biggl(C_n^{ss} \bigl[\sinh\alpha_2 
+ \cosh\alpha_2 C_n^{+}\bigr] + C_n^{cc} \sinh\alpha_1 + \cosh\alpha_1 C_n^{sc}\biggr)  
 - \bigl[C_n^{ss} C_{n'}^{+} + C_{n'}^{sc} \bigr]  [\mm_m]_{n',n}  \biggr].
\end{align*}
\end{widetext}
A numerical diagonalization of the truncated matrix $\MM_m$ allows one
to approximate the Steklov eigenvalues $\mu_{m,n}$, enumerated by $n$,
whereas the left eigenvectors determine the associated eigenfunctions.
Note that the numerical computation becomes unstable in the case
$\alpha_1 > 0$ because $C_n^{cc} - C_n^{ss}$ is very small for large
$n$.  An alternative form of the linear system may be needed in this
case.


\begin{thebibliography}{99}


\bibitem{House}			J. E. House, 
				{\it Principles of chemical kinetics}, 2nd Ed.
				(Academic press: Amsterdam, 2007).

\bibitem{Murrey}		J. D. Murrey,
				{\it Mathematical Biology II: Spatial Models and Biomedical Applications}, 3rd Ed.
				(Springer: Berlin, Germany, 2003).

\bibitem{Lindenberg}		K. Lindenberg, R. Metzler, and G. Oshanin (Eds.)
				{\em Chemical Kinetics: Beyond the Textbook}
				(World Scientific: New Jersey, 2019).

\bibitem{Grebenkov}		D. S. Grebenkov, R. Metzler, and G. Oshanin (Eds), 
				{\it Target Search Problems} 
				(Springer: Cham, Switzerland, 2024). 



\bibitem{Smoluchowski18} 	M. von Smoluchowski, 
				Versuch einer Mathematischen Theorie der Koagulations Kinetic Kolloider Lousungen. 
				Z. Phys. Chem. {\bf 92U}, 129-168 (1918).


\bibitem{North66}		A. M. North,  
				Diffusion-controlled reactions,
				Q. Rev. Chem. Soc. {\bf 20}, 421-440 (1966).

\bibitem{Wilemski73}		G. Wilemski and M. Fixman, 
				General theory of diffusion-controlled reactions, 
				J. Chem. Phys. {\bf 58}, 4009-4019 (1973).

\bibitem{Calef83}		D. F. Calef and J. M. Deutch, 
				Diffusion-Controlled Reactions,
				Ann. Rev. Phys. Chem. {\bf 34}, 493-524 (1983).

\bibitem{Berg85}		O. G. Berg and P. H.  von Hippel,  
				Diffusion-Controlled Macromolecular Interactions, 
				Ann. Rev. Biophys. Biophys. Chem. {\bf 14}, 131-160 (1985).

\bibitem{Rice85}		S. Rice, 
				{\em Diffusion-Limited Reactions}
				(Elsevier: Amsterdam, The Netherlands, 1985).

\bibitem{Weiss86} 		G. H. Weiss,  
				Overview of theoretical models for reaction rates,
				J. Stat. Phys. {\bf 42}, 3-36 (1986).

\bibitem{Szabo89} 		A. Szabo,  
				Theory of diffusion-influenced fluorescence quenching,
				J. Phys. Chem. {\bf 93}, 6929-6939 (1989).

\bibitem{Zhou10} 		H.-X. Zhou,
				Rate theories for biologists,
				Quart. Rev. Biophys. {\bf 43}, 219-293 (2010).

\bibitem{Grebenkov23}		D. S. Grebenkov, 
				Diffusion-Controlled Reactions: An Overview, 
				Molecules {\bf 28}, 7570 (2023).





\bibitem{Collins49}		F. C. Collins and G. E. Kimball,
				Diffusion-controlled reaction rates,
				J. Coll. Sci. {\bf 4}, 425 (1949).


\bibitem{Sano79}		H. Sano and M. Tachiya,
				Partially diffusion-controlled recombination,
				J. Chem. Phys. {\bf 71}, 1276-1282 (1979).

\bibitem{Brownstein79}		K. R. Brownstein and C. E. Tarr, 
				Importance of Classical Diffusion in NMR Studies of Water in Biological Cells, 
				Phys. Rev. A {\bf 19}, 2446-2453 (1979).

\bibitem{Sano81}		H. Sano and  M. Tachiya,
				Theory of diffusion-controlled reactions on spherical surfaces and 
				its application to reactions on micellar surfaces,
				J. Chem. Phys. {\bf 75}, 2870-2878 (1981).

\bibitem{Shoup82}		D. Shoup and A. Szabo, 
				Role of diffusion in ligand binding to macromolecules and cell-bound receptors'', 
				Biophys. J. {\bf 40}, 33-39 (1982).

\bibitem{Zwanzig90}		R. Zwanzig, 
				Diffusion-controlled ligand binding to spheres partially covered by receptors: an effective medium treatment,
				Proc. Natl. Acad. Sci. USA {\bf 87}, 5856 (1990).

\bibitem{Powles92}		J. G. Powles, M. J. D. Mallett, G. Rickayzen, and W. A. B. Evans, 
				Exact analytic solutions for diffusion impeded by an infinite array of partially 
				permeable barriers, 
				Proc. R. Soc. London A {\bf 436}, 391-403 (1992).


\bibitem{Sapoval94}		B. Sapoval,
				General Formulation of Laplacian Transfer Across Irregular Surfaces,
				Phys. Rev. Lett. {\bf 73}, 3314 (1994).

\bibitem{Berezhkovskii04}	A. Berezhkovskii, Y. Makhnovskii, M. Monine, V. Zitserman, and S. Shvartsman, 
				Boundary homogenization for trapping by patchy surfaces,
				J. Chem. Phys. {\bf 121}, 11390 (2004).



\bibitem{Traytak07}		S. D. Traytak and W. Price,
				Exact solution for anisotropic diffusion-controlled reactions with partially reflecting conditions,
				J. Chem. Phys. {\bf 127}, 184508 (2007).

\bibitem{Bressloff08}		P. C. Bressloff, B. A. Earnshaw, and M. J. Ward,
				Diffusion of protein receptors on a cylindrical dendritic membrane with partially absorbing traps,
				SIAM J. Appl. Math. {\bf 68}, 1223-1246 (2008).


\bibitem{Singer08b}		A. Singer, Z. Schuss, A. Osipov, and D. Holcman,
				Partially reflected diffusion,
				SIAM J. Appl. Math. {\bf 68}, 844-868 (2008).

\bibitem{Lawley15}		S. D. Lawley and J. P. Keener,
				A New Derivation of Robin Boundary Conditions through Homogenization of a Stochastically Switching Boundary,
				SIAM J. Appl. Dyn. Sys. {\bf 14}, 1845-1867 (2015).


\bibitem{Guerin21}		T. Gu\'erin, M. Dolgushev, O. B\'enichou, and R. Voituriez,
				Universal kinetics of imperfect reactions in confinement,
				Commun. Chem. {\bf 4}, 157 (2021).

\bibitem{Piazza22}		F. Piazza,
				The physics of boundary conditions in reaction-diffusion problems,
				J. Chem. Phys. {\bf 157}, 234110 (2022).

\bibitem{Guerin23}		T. Gu\'erin, M. Dolgushev, O. B\'enichou, and R. Voituriez,
				Imperfect narrow escape problem,
				Phys. Rev. E {\bf 107}, 034134 (2023).

\bibitem{Cengiz24}		A. Cengiz and S. D. Lawley,
				Narrow escape with imperfect reactions,
				Phys. Rev. E {\bf 110}, 054127 (2024).




\bibitem{Grebenkov20}		D. S. Grebenkov,
				Paradigm shift in diffusion-mediated surface phenomena,
				Phys. Rev. Lett. {\bf 125}, 078102 (2020).




\bibitem{Piazza19}		F. Piazza and D. S. Grebenkov,
				Diffusion-controlled reaction rate on non-spherical partially absorbing axisymmetric surfaces,
				Phys. Chem. Chem. Phys. {\bf 21}, 25896-25906 (2019).



\bibitem{Smythe}		W. R. Smythe,
				Static and dynamic electricity, 3rd Ed.
				(Taylor \& Francis, 1989).




\bibitem{Samson77}  		R. Samson and J. M. Deutch, 
				Exact solution for the diffusion controlled rate into a pair of reacting sinks,
				J. Chem. Phys. {\bf 67}, 847 (1977).

\bibitem{Zoia98}  		G. Zoia and W. Strieder, 
				Competitive diffusion into two sinks with a finite surface reaction coefficient,
				J. Chem. Phys. {\bf 108}, 3114 (1998).

\bibitem{Strieder00}  		W. Strieder and S. Saddawi, 
				Alternative solution for diffusion to two spheres with first-order surface reaction,
				J. Chem. Phys. {\bf 113}, 10818 (2000).

\bibitem{Piazza05} 		F. Piazza, P. De Los Rios, D. Fanelli, L. Bongini, and U. Skoglund, 
				Anti-cooperativity in diffusion-controlled reactions with pairs of anisotropic domains: 
				a model for the antigen-antibody encounter,
				Eur. Biophys. J. {\bf 34}, 899 (2005).

\bibitem{Bluett06}		V. M. Bluett and N. J. B. Green, 
				Competitive diffusion-influenced reaction of a reactive particle with two static sinks,
				J. Phys. Chem. A {\bf 110}, 4738 (2006).




\bibitem{Popescu11} 		M. N. Popescu, M. Tasinkevych and S. Dietrich, 
				Pulling and pushing a cargo with a catalytically active carrier,
				EPL {\bf 95}, 28004 (2011).

\bibitem{Kapral15}  		S. Y. Reigh and R. Kapral, 
				Catalytic dimer nanomotors: continuum theory and microscopic dynamics,
				Soft Matter {\bf 11}, 3149 (2015).

\bibitem{Michelin17} 		S. Michelin and E. Lauga,
				Geometric tuning of self-propulsion for Janus catalytic particles, 
				Sci. Rep. {\bf 7}, 42264 (2017).









\bibitem{Zheng89}		L. H. Zheng and Y. C. Chiew,
				Computer simulation of diffusion-controlled reactions in dispersions of spherical sinks,
				J. Chem. Phys. {\bf 90}, 322 (1989).

\bibitem{Tsao01}		H.-K. Tsao, S.-Y. Lu, and C.-Y. Tseng,
				Rate of diffusion-limited reactions in a cluster of spherical sinks,
				J. Chem. Phys. {\bf 115}, 3827 (2001).

\bibitem{Kansal02}		A. R. Kansal and S. Torquato,
				Prediction of trapping rates in mixtures of partially absorbing spheres,
				J. Chem. Phys. {\bf 116}, 10589 (2002).



\bibitem{Eun13}			C. Eun, P. M. Kekenes-Huskey, and J. A. McCammon,
				Influence of neighboring reactive particles on diffusion-limited reactions, 
				J. Chem. Phys. {\bf 139}, 044117 (2013).



\bibitem{Galanti16}		M. Galanti, D. Fanelli, S. Traytak, and F. Piazza,
				Theory of diffusion-influenced reactions in complex geometries,
				Phys. Chem. Chem. Phys. {\bf 18}, 15950-15954 (2016).

\bibitem{Galanti16b}		M. Galanti, D. Fanelli, S. Angioletti-Uberti, M. Ballauff, J. Dzubiella, and F. Piazza,
				Reaction rate of a composite core-shell nanoreactor with multiple nanocatalysts,
				Phys. Chem. Chem. Phys. {\bf 18}, 20758-20767 (2016).

\bibitem{Traytak18}		S. D. Traytak and D. S. Grebenkov, 
				Diffusion-influenced reaction rates for active ``sphere-prolate spheroid'' pairs and Janus dimers, 
				J. Chem. Phys. {\bf 148}, 024107 (2018).

\bibitem{Grebenkov19f}		D. S. Grebenkov and S. Traytak, 
				Semi-analytical computation of Laplacian Green functions in three-dimensional 
				domains with disconnected spherical boundaries, 
				J. Comput. Phys. {\bf 379}, 91-117 (2019).

\bibitem{Grebenkov20f}		D. S. Grebenkov, 
				Diffusion toward non-overlapping partially reactive spherical traps: 
				fresh insights onto classic problems, 
				J. Chem. Phys. {\bf 152}, 244108 (2020).


\bibitem{Grebenkov06b}		D. S. Grebenkov, M. Filoche, and B. Sapoval,
				Mathematical Basis for a General Theory of Laplacian Transport towards Irregular Interfaces,
				Phys. Rev E {\bf 73}, 021103 (2006).

\bibitem{Filoche08}		M. Filoche and D. S. Grebenkov,
				The toposcopy, a new tool to probe the geometry of an irregular interface by measuring its transfer impedance,
				Euro. Phys. Lett. {\bf 81}, 40008 (2008).










\bibitem{Auchmuty14}		G. Auchmuty and Q. Han,
				Representations of Solutions of Laplacian Boundary Value Problems on Exterior Regions,
				Appl. Math. Optim. {\bf 69}, 21-45 (2014).  

\bibitem{Arendt15}		W. Arendt and A. F. M. ter Elst,
				The Dirichlet-to-Neumann Operator on Exterior Domains,
				Potential Anal. {\bf 43}, 313-340 (2015).

\bibitem{Bundrock25}		L. Bundrock, A. Girouard, D. S. Grebenkov, M. Levitin, and I. Polterovich,
				Exterior Steklov problems
				(in preparation).



\bibitem{Levitin}		M. Levitin, D. Mangoubi, and I. Polterovich,
				{\it Topics in Spectral Geometry}
				(Graduate Studies in Mathematics, vol. 237; American Mathematical Society, 2023).

\bibitem{Girouard17}		A. Girouard and I. Polterovich,
				Spectral geometry of the Steklov problem, 
				J. Spectr. Th. {\bf 7}, 321-359 (2017).

\bibitem{Colbois24}		B. Colbois, A. Girouard, C. Gordon, and D. Sher,
				Some recent developments on the Steklov eigenvalue problem,
				Rev. Mat. Complut. {\bf 37}, 1-161 (2024).




\bibitem{Henrici70}		P. Henrici, B. A. Troesch, L. Wuytack,
				Sloshing frequencies for a half-space with circular or strip-like aperture,
				Z. Angew. Math. Phys. {\bf 21}, 285-318 (1970).

\bibitem{Troesch72}		B. A. Troesch and H. R. Troesch,
				A remark on the sloshing frequencies for a half-space,
				J. Appl. Math. Phys. {\bf 23}, 703 (1972).

\bibitem{Miles72}		J. W. Miles,
				On the eigenvalue problem for fluid sloshing in a half-space,
				Z. Angew. Math. Phys. {\bf 23}, 861-868 (1972).

\bibitem{Fox83}			D. W. Fox and J. R. Kuttler,
				Sloshing frequencies,
				Z. Angew. Math. Phys. {\bf 34}, 668-696 (1983).

\bibitem{Kozlov04}		V. Kozlov and N. Kuznetsov,
				The ice-fishing problem: the fundamental sloshing frequency versus geometry of holes,
				Math. Methods Appl. Sci. {\bf 27}, 289-312 (2004).



\bibitem{Auchmuty04}		G. Auchmuty,
				Steklov eigenproblems and the representation of solutions of 
				elliptic boundary value problems,
				Numer. Funct. Anal. Optim. {\bf 25}, 321-348 (2005).

\bibitem{Auchmuty13}		G. Auchmuty and Q. Han,
				Spectral representations of solutions of linear elliptic equations on exterior regions,
				J. Math. Anal. Appl. {\bf 398}, 1-10 (2013).

\bibitem{Auchmuty15}		G. Auchmuty and M. Cho, 
				Boundary integrals and approximations of harmonic functions, 
				Numer. Funct. Anal. Optim. {\bf 36}, 687-703 (2015).

\bibitem{Auchmuty18}		G. Auchmuty,
				Steklov Representations of Green's Functions for Laplacian Boundary Value Problems,
				Appl. Math. Optim. {\bf 77} (2018).


\bibitem{Grebenkov21a}		D. S. Grebenkov, 
				Statistics of boundary encounters by a particle diffusing outside a compact planar domain, 
				J. Phys. A.: Math. Theor. {\bf 54}, 015003 (2021).

\bibitem{Grebenkov25}		D. S. Grebenkov and A. Chaigneau,
				The Steklov problem for exterior domains: asymptotic behavior and applications
				(submitted; see preprint on ArXiv 2407.09864).




\bibitem{Sapoval02}		B. Sapoval, M. Filoche, and E. Weibel,
				Smaller is better -- but not too small: A physical scale for the design of the mammalian pulmonary acinus,
				Proc. Nat. Acad. Sci. USA {\bf 99}, 10411 (2002).

\bibitem{Felici03}		M. Felici, M. Filoche, and B. Sapoval,
				Diffusional screening in the human pulmonary acinus,
				J. Appl. Physiol. {\bf 94}, 2010 (2003).

\bibitem{Felici05}		M. Felici, M. Filoche, C. Straus, T. Similowski, and B. Sapoval,
				Diffusional screening in real 3D human acini: a theoretical study,
				Resp. Physiol. Neurobiol. {\bf 145}, 279 (2005).


\bibitem{Grebenkov15}		D. S. Grebenkov,
				Analytical representations of the spread harmonic measure density,
				Phys. Rev. E {\bf 91}, 052108 (2015).



\bibitem{Belevitch83}		V. Belevitch and J. Boersma, 
				Some electrical problems for a torus,
				Philips J. Res. {\bf 38}, 79-137 (1983).



\bibitem{Grebenkov17a}		D. S. Grebenkov and G. Oshanin, 
				Diffusive escape through a narrow opening: new insights into a classic problem,
				Phys. Chem. Chem. Phys. {\bf 19}, 2723-2739 (2017).

\bibitem{Grebenkov18}		D. S. Grebenkov, R. Metzler, and G. Oshanin, 
				Strong defocusing of molecular reaction times results from an interplay of geometry and reaction control, 
				Commun. Chem. {\bf 1}, 96 (2018).







\bibitem{Traytak92} 		S. D. Traytak,
				The diffusive interaction in diffusion-limited reactions: the steady-state case,
				Chem. Phys. Lett. {\bf 197}, 247-254 (1992).



\bibitem{Witten81}		T. A. Witten and L. M. Sander,
				Diffusion-Limited Aggregation, a Kinetic Critical Phenomenon,
				Phys. Rev. Lett. {\bf 47}, 1400-1403 (1981).

\bibitem{Vicsek}		T. Vicsek, 
				{\it Fractal Growth Phenomena}, 2nd ed. 
				(Singapore, World Scientific, 1992).

\bibitem{Banks}			R. B. Banks, 
				{\it Growth and Diffusion Phenomena: Mathematical Frameworks and Applications} 
				(Springer Science \& Business Media, 1994).

\bibitem{Krapivsky}		P. L. Krapivsky, S. Redner, and E. Ben-Naim, 
				{\it A kinetic view of statistical physics} 
				(Cambridge University Press, Cambridge, 2010).



\bibitem{Chaigneau22}		A. Chaigneau and D. S. Grebenkov,
				First-passage times to anisotropic partially reactive targets,
				Phys. Rev. E {\bf 105}, 054146 (2022).


\bibitem{Grebenkov24}		D. S. Grebenkov, 
				Spectral properties of the Dirichlet-to-Neumann operator for spheroids, 
				Phys. Rev. E {\bf 109}, 055306 (2024). 




\bibitem{Grebenkov19}		D. S. Grebenkov,
				Spectral theory of imperfect diffusion-controlled reactions on heterogeneous catalytic surfaces,
				J. Chem. Phys. {\bf 151}, 104108 (2019).













\bibitem{Grebenkov20c}		D. S. Grebenkov, 
				Surface Hopping Propagator: An Alternative Approach to Diffusion-Influenced Reactions, 
				Phys. Rev. E {\bf 102}, 032125 (2020).


\bibitem{Grebenkov22a}		D. S. Grebenkov, 
				An encounter-based approach for restricted diffusion with a gradient drift, 
				J. Phys. A: Math. Theor. {\bf 55}, 045203 (2022). 


\bibitem{Bressloff22d}		P. C. Bressloff,
				Diffusion-mediated surface reactions and stochastic resetting,
				J. Phys. A: Math. Theor. {\bf 55}, 275002 (2022).


\bibitem{Bressloff22b}		P. C. Bressloff,
				Narrow capture problem: an encounter-based approach to partially reactive targets,
				Phys. Rev. E {\bf 105}, 034141 (2022).




\bibitem{Bressloff22a}		P. C. Bressloff,
				Diffusion-mediated absorption by partially-reactive targets: 
				Brownian functionals and generalized propagators,
				J. Phys. A: Math. Theor. {\bf 55}, 205001 (2022).

\bibitem{Bressloff22c}		P. C. Bressloff, 
				A probabilistic model of diffusion through a semipermeable barrier,
				Proc. Roy. Soc. A {\bf 478}, 20220615 (2022).



\bibitem{Bressloff23a}		P. C. Bressloff,
				Renewal equation for single-particle diffusion through a semipermeable interface,
				Phys. Rev. E. {\bf 107}, 014110 (2023).

\bibitem{Bressloff23b}		P. C. Bressloff,
				Renewal equations for single-particle diffusion in multilayered media,
				SIAM J. Appl. Math. {\bf 83}, 1518-1545 (2023).

\bibitem{Grebenkov19c}		D. S. Grebenkov,
				Probability distribution of the boundary local time of reflected Brownian motion in Euclidean domains,
				Phys. Rev. E {\bf 100}, 062110 (2019).


\bibitem{Grebenkov22d}		D. S. Grebenkov, 
				Statistics of diffusive encounters with a small target: Three complementary approaches, 
				J. Stat. Mech. 083205 (2022). 






\bibitem{Morse}			P. Morse and H. Feshbach,
				{\it Methods of Theoretical Physics}, Part 2
				(McGraw-Hill Book Company, New York, 1953).



\bibitem{NIST_Pnu}		NIST 14.12.6, \url{https://dlmf.nist.gov/14.12}

\bibitem{Abramowitz}		M. Abramowitz and I. A. Stegun,
				{\it Handbook of Mathematical Functions}
				(Dover Publisher, New York, 1965).

\bibitem{NIST14.5.25}		NIST 14.5.25, \url{https://dlmf.nist.gov/14.5}




\bibitem{Chiado20}		V. Chiad\`o Piat and S. A. Nazarov,
				Steklov spectral problems in a set with a thin toroidal hole,
				PDE Appl. Math. {\bf 1}, 100007 (2020).


\bibitem{Brisson22}		J. Brisson,
				Tubular Excision and Steklov eigenvalues,
				J. Geom. Anal. {\bf 32}, 166 (2022).

























\end{thebibliography}
\end{document}